\documentclass[12pt]{article}
\usepackage{amsmath}
\usepackage[authoryear]{natbib}
\usepackage{latexsym, epsfig, amssymb, amsmath, amsthm, graphicx, mathrsfs}
\usepackage[dvips]{color}
\usepackage[OT1]{fontenc}
\usepackage{anysize}
\def\singlespace{\def\baselinestretch{1}\@normalsize}
\renewcommand{\baselinestretch}{1.5}
\renewcommand{\baselinestretch}{1.18}
\marginsize{1.2in}{0.9in}{0.5in}{1.3in} %
\makeatletter

\usepackage[inline]{trackchanges}

\newtheorem{assumption}{Condition}%[section]
\newtheorem{lemma}{Lemma}%[section]
\newtheorem{proposition}{Proposition}%[section]
\newtheorem{theorem}{Theorem}%[section]
\newtheorem{corollary}{Corollary}%[section]
\renewcommand{\theequation}{%\thesection.
\arabic{equation}%
}
\renewcommand{\thefootnote}{\fnsymbol{footnote}}
\makeatother

\newcommand{\bb}{\mbox{\bf b}}

\newcommand{\bv}{\mbox{\bf v}}

\newcommand{\bx}{\mbox{\bf x}}

\newcommand{\bB}{\mbox{\bf B}}

\newcommand{\bH}{\mbox{\bf H}}

\newcommand{\bP}{\mbox{\bf P}}

\newcommand{\bW}{\mbox{\bf W}}
\newcommand{\bX}{\mbox{\bf X}}

\newcommand{\bY}{\mbox{\bf Y}}

\newcommand{\bone}{\mbox{\bf 1}}

\newcommand{\bbeta}{\mbox{\boldmath $\beta$}}

\newcommand{\bmu}{\mbox{\boldmath $\mu$}}

\newcommand{\hbbeta}{\widehat\bbeta}

\newcommand{\tbbeta}{\widetilde\bbeta}

\newcommand{\hbeta}{\widehat\beta}

\newcommand{\hbmu}{\widehat\bmu}

\newcommand{\sbbeta}{\mbox{\scriptsize \boldmath $\beta$}}

\newcommand{\veps}{\varepsilon}
\newcommand{\tr}{\mathrm{tr}}

\newcommand{\sgn}{\mathrm{sgn}}
\newcommand{\supp}{\mathrm{supp}}

\newcommand{\GIC}{\mbox{GIC}}

\def\t{^T}

\renewcommand{\theequation}{\thesection.\arabic{equation}}
\begin{document}

\DeclareGraphicsExtensions{.pdf,.gif,.jpg}

\title{Tuning Parameter Selection in High-Dimensional Penalized Likelihood}

\date{}
\author{{\sc Yingying Fan and Cheng Yong Tang
\thanks{Marshall
 School of Business, University of Southern California, and Department of Statistics and Applied Probability, National University of Singapore. We thank the Editor, the AE,  and two referees for their insightful comments and constructive suggestions that have greatly improved the presentation of the paper. Fan's work was
  partially supportedby NSF CAREER Award DMS-1150318 and Grant DMS-0906784, and 2010 Zumberge Individual Award from
USC's James H. Zumberge Faculty Research and Innovation
Fund. Tang acknowledges support from National University of Singapore Academic Research Grants and a Research Grant from Risk Management Institute, National University of Singapore.  }
}}
\maketitle
%\vspace{-2cm}

\begin{abstract}

Determining how to appropriately select the tuning parameter  is essential  in penalized likelihood methods for high-dimensional data analysis. We examine this problem in the setting of penalized likelihood methods for generalized linear models, where the dimensionality of covariates $p$   is allowed to increase exponentially with the sample size $n$.  We propose to select the tuning parameter
by optimizing the generalized information criterion (GIC) with an appropriate model complexity penalty.
To ensure  that we consistently identify the true model, a range for the model complexity penalty is identified in GIC. We find that this model complexity penalty should diverge at the rate of some power of $\log p$ depending on the tail probability behavior of the response variables.  This reveals that using the AIC or BIC to select the tuning parameter may not be adequate for consistently identifying the true model.  Based on our theoretical study, we  propose a uniform choice of the model complexity penalty and show that the proposed approach consistently identifies the true model among candidate models with asymptotic probability one.   We justify the performance of the proposed procedure by numerical simulations and a gene expression data analysis.

\end{abstract}


\section{Introduction}

Various types of high-dimensional data are encountered in multiple disciplines when solving practical problems, for example,  gene expression data for disease classifications \citep{Golubetal_1999_SCI}, financial market data for portfolio construction and assessment \citep{JM_2003_JOF}, and spatial earthquake data for geographical analysis \citep{vanderHilstetal_2007_SCI},  among many others.    To meet the challenges in analyzing high-dimensional data, penalized likelihood methods have been extensively studied; see \cite{Hastieetal_2009_book} and  \cite{FanLv_2009_Sinica} for overviews among a large amount of recent literature.

Though demonstrated effective in analyzing high-dimensional data, the performance of penalized likelihood methods depends on the choice of the tuning parameters, which controls the trade-off between the bias and variance in resulting estimators \citep{Hastieetal_2009_book, FanLv_2009_Sinica}.
Generally speaking, the optimal properties of those penalized likelihood methods require certain specifications of the optimal tuning parameters \citep{FanLv_2009_Sinica}.  On the other hand, theoretically quantified optimal tuning parameters are not practically feasible, because they are valid only asymptotically and usually depend on unknown nuisance parameters in the true model.  Therefore, in practical implementations, penalized likelihood methods are usually applied with a sequence of tuning parameters resulting in a corresponding collection of models.  Then, selecting an appropriate model and equivalently the corresponding tuning parameter becomes an important question of interest, both theoretically and practically.

Traditionally in model selection, cross-validation  and information criteria -- including AIC \citep{AIC1973} and BIC \citep{Schwarz_1978} -- are widely applied.   A generalized information criterion \citep{Nishii_1984_AOS} is constructed as follows:
\begin{equation}\label{e019}
\text{measure of model fitting} + a_n \times \text{ measure of model complexity},
\end{equation}
where $a_n$ is some positive sequence that depends only on the sample size $n$ and that controls the penalty on model complexity.   The rationale of the information criteria for model selection is that the true model can uniquely optimize the information criterion (\ref{e019}) by appropriately choosing $a_n$.    Hence, the choice of $a_n$ becomes crucial for effectively identifying the true model.  The minus log likelihood is commonly used as a  measure of the model fitting, and $a_n$ is $2$ and $\log n$  in the AIC and BIC respectively. It is known that the BIC can identify the true model consistently in linear regression with fixed dimensional covariates, while the AIC may fail due to overfitting  \citep{Shao_1997_Sinica}. Meanwhile, cross-validation is shown asymptotically equivalent to the AIC \citep{Yang_2005_Bioka} so that they behave similarly.

When applying penalized likelihood methods, existing model selection criteria are naturally incorporated to select the tuning parameter.  Analogously to those results for model selection,  \cite{Wangetal_2007_Bioka} showed that the tuning parameter selected by the BIC criterion can identify the true model consistently  for the SCAD approach in \cite{FanLi_2001_JASA}, while the AIC and cross-validation may fail to play such a role (see also \citeauthor{ZLT2010}, \citeyear{ZLT2010}).  However, those studies on tuning parameter selection for penalized likelihood methods are mainly for fixed dimensionality.
\cite{Wangh2009} recently considered the tuning parameter selection in the setting of linear regression with diverging dimensionality and showed that a modified BIC criterion continues to work for tuning parameter selection.  However, their analysis is confined to the penalized least-squares method, and the dimensionality $p$ of covariates is not allowed to exceed the sample size $n$. We also refer to \cite{ChenChen_2008_Bioka} for a recent study on an extended BIC and its property for Gaussian linear models, and \cite{WangZhu_2011_JMVA} for tuning parameter selection in high-dimensional penalized least-squares.

The current trend  of  high-dimensional data analysis poses new challenges for tuning parameter selection.  To the best of our knowledge, there is no existing work accommodating tuning parameter selection for general penalized likelihood methods when the dimensionality $p$ grows exponentially with the sample size $n$, i.e. $\log p = O(n^\kappa)$ for some $\kappa>0$.  The problem is challenging in a few aspects.  First,  note that there are generally no explicit forms of the maximum likelihood estimates for models other than the linear regression model, which makes it more difficult to characterize the asymptotic performance of the first part of (\ref{e019}). Second, the exponentially growing dimensionality $p$ induces a huge number of candidate models. One may reasonably conjecture that the true model may be differentiated from a specific candidate model with  probability tending to 1 as $n\to\infty$. However, the probability that the true model is not dominated by any of the candidate models may not be straightforward to calculate, and an inappropriate choice of $a_n$ in (\ref{e019}) may even cause the model selection consistency to fail.

%Motivated by recent developments of penalization methods,
We explore in this paper the tuning parameter selection for penalized generalized linear regression, with the penalized Gaussian linear regression as a special case,  in which the dimensionality $p$ is allowed to increase exponentially fast with the sample size $n$.
We systematically examine the generalized information criterion, and our analysis reveals the connections among the model complexity penalty $a_n$, the data dimensionality $p$,  and the tail probability distribution of the response variables, for consistently identifying the true model.
 Subsequently,  we identify a range of $a_n$  such that the tuning parameter selected by optimizing the generalized information criterion can achieve model selection consistency. We find that when $p$ grows polynomially with sample size $n$, the modified BIC criteria \citep{Wangh2009} can still be successful in tuning parameter selection. But when $p$ grows exponentially with sample size $n$, $a_n$ should diverge with some power of $\log p$, where the power depends on the tail distribution of response variables. This produces  a phase diagram of how the model complexity penalty should adapt to the growth of sample size $n$ and dimensionality $p$.
Our theoretical investigations, numerical implementations by simulations,  and a data analysis illustrate that the proposed approach can be effectively and conveniently applied in practice.
As demonstrated in Figure \ref{fig3} for analyzing a gene expression data-set, we find that a single gene identified by the proposed approach can be very informative in predictively differentiating between two types of leukemia patients.

  %More specifically, we find that the heavier tail distribution of the response variable, the larger value that the model complexity penalty $a_n$ should take to achieve model selection consistency.
 %Based on our theoretical study, we identify the choice of $a_n$ which can lead to model selection consistency. %We discover that
 %When $p$ grows polynomially with sample size $n$, the modified BIC criteria \citep{Wangh2009} can still success in tuning parameter selection. But
 %When $p$ grows exponentially with sample size $n$,

 The rest of this paper is organized as follows.  In Section \ref{sec:modelsetting}, we outline the problem and define the model selection criterion GIC. To study GIC, we first investigate the asymptotic property of a proxy of GIC in Section \ref{sec:proxyGIC}, and we summarize the main result of the paper  in Section \ref{sec:tune}.
 Section \ref{sec:numeric} demonstrates the proposed approach via numerical examples of simulations and gene expression data analysis, and Section \ref{sec:results} contains the technical conditions and some intermediate results. The technical proofs are contained in the Appendix.

%\section{Model Setting and Technical Conditions}
\setcounter{equation}{0}

\section{Penalized Generalized Linear Regression and Tuning Parameter Selection}\label{sec:modelsetting}

Let $\{(\bx_i, Y_i)\}_{i=1}^n$ %(\bx_2, Y_2), \cdots, (\bx_n, Y_n)$
be independent data with the scalar response variable   $Y_i$  and %from the $i$th observation, and
the corresponding $p$-dimensional covariate vector $\bx_i$  for the $i$th observation. %Suppose that the underlying true model is
We consider the generalized linear model \citep{GLM_1989_MN} with  the conditional density function of $Y_i$ given $\bx_i$
\begin{align}\label{e016}
f_i(y_i; \theta_i, \phi) = \exp\{y_i\theta_i - b(\theta_i) + c(y_i, \phi)\},
\end{align}
where  $\theta_i = \bx_i\t\bbeta$ is the canonical parameter with $\bbeta$ a $p$-dimensional regression coefficient,  $b(\cdot)$ and $c(\cdot,\cdot)$ are some suitably chosen known functions, $E[Y_i|\bx_i]=\mu_i=b'(\theta_i)$, $g(\mu_i)=\theta_i$ is the link function, and $\phi$ is a known scale parameter. Thus, the log-likelihood function for $\bbeta$ is given by
\begin{align}
\ell_n(\bbeta)& = %\ell_n(\bmu; \bY) = \ell_n(\btheta)
%= \sum_{i=1}^n\log f_i(Y_i; \theta_i, \phi)=
\sum_{i=1}^n \{Y_i\bx_i^T\bbeta - b(\bx_i^T\bbeta) + c(Y_i, \phi)\}. \label{eq:lhood}
\end{align}

The dimensionality $p$ in our study is allowed to increase with sample size $n$ exponentially fast -- i.e., $\log p=O(n^\kappa)$ for some $\kappa>0$. To enhance  the model fitting accuracy and ensure the model identifiability, it is commonly assumed that the true population parameter $\bbeta_0$ is sparse,  with only a small fraction of nonzeros \citep{tibshirani2,FanLi_2001_JASA}. Let $\alpha_0 = \text{supp}(\bbeta_0)$ be the support of the true model consisting of indices of all non-zero components in $\bbeta_0$, and let $s_n=|\alpha_0|$ be the number of true covariates, which may increase with $n$  and which satisfies $s_n=o(n)$.
To ease the presentation, we suppress the dependence of $s_n$ on $n$ whenever there is no confusion.
By using compact notations,  we write  $\bY = (Y_1 ,\cdots, Y_n)^T$ as the $n$-vector of response, $\bX = (\bx_1, \cdots, \bx_n)^T = (\tilde\bx_1, \cdots, \tilde\bx_p)$ as the $n\times p$ fixed design matrix, and  $\bmu=\bb'(\bX\bbeta)=(b'(\bx_1^T\bbeta),\dots, b'(\bx_n^T\bbeta))^T$ as the mean vector.  We standardize each column of $\bX$ so that $\|\tilde\bx_j\|_2=\sqrt{n}$ for $j=1,\cdots, p$.

In practice, the true parameter $\bbeta_0$ is unknown and needs to be estimated from data.
Penalized likelihood methods have attracted substantial attention recently for simultaneously selecting and estimating the unknown parameters.
  The penalized maximum likelihood estimator (MLE) is broadly defined as
\begin{align}\label{e001}
\hbbeta^{\lambda_n}={\arg\max_{{\scriptsize\bbeta}\in \mathbf{R}^p}}
\Big\{ \ell_n(\bbeta) - n\sum_{j=1}^{p}p_{\lambda_n}(|\beta_j|)\big\},
\end{align}
where $p_{\lambda_n}(\cdot)$ is some penalty function with tuning parameter $\lambda_n\geq 0$. For simplicity, we suppress the dependence of $\lambda_n$ on $n$ and write it as $\lambda$ when there  is no confusion. Let $\alpha_\lambda=\text{supp}(\hbbeta^{\lambda})$ be the model identified by the penalized likelihood method with tuning parameter $\lambda$.

For the penalized likelihood method to successfully identify the underlying true model and enjoy desirable  properties, it is critically important to choose an appropriate tuning parameter $\lambda$. Intuitively, a too large (small) tuning parameter imposes an excessive (inadequate) penalty on the magnitude of the parameter so that the support of  $\hbbeta^\lambda$ is different from that of the true model $\alpha_0$.
Clearly, a meaningful discussion of tuning parameter selection in (\ref{e001})
requires the existence of a $\lambda_0$ such that $\alpha_{\lambda_0} = \alpha_0$,
 %Existence of such a $\lambda_0$
 which has been established in various model settings when different penalty functions are used; see, for example,  \cite{ZhaoYu_2006_JMLR}, \cite{lv1},  and  \cite{lv2}.  %for conditions ensuring the existence of such $\lambda_0$.

To identify  the $\lambda_0$ that leads to the true model $\alpha_0$, we propose to use the generalized information criterion (GIC):
\begin{align}\label{e002}
\GIC_{a_n}(\lambda) = \frac{1}{n}\left\{D(\hbmu_{\lambda}; \bY)+ a_n |\alpha_\lambda|\right\},
\end{align}
where $a_n$ is a positive sequence depending only on $n$ and $D( \hbmu_\lambda; \bY)$ is the scaled deviation measure defined as the scaled log-likelihood ratio of the saturated model and the candidate model with parameter $\hbbeta^{\lambda}$;  i.e.,
\begin{equation}\label{e023}
D( \hbmu_\lambda; \bY) = 2\{\ell_n(\bY;\bY)-\ell_n(\hbmu_\lambda;\bY)\}
\end{equation}
with $\ell_n(\bmu;\bY)$ the log-likelihood function (\ref{eq:lhood}) expressed as  a function of $\bmu$ and $\bY$, and $\hbmu_\lambda = \bb'(\bX\hbbeta^{\lambda})$. %=(b'(\bx_1\t\hbbeta^{\lambda}), \cdots, b'(\bx_n\t\hbbeta^{\lambda}))\t$, and $b'(\cdot)$ the first order derivative of $b(\cdot)$.
The scaled deviation measure is used to evaluate the goodness-of-fit. It reduces to the sum of squared residuals in Gaussian linear regression. The second component in the definition of GIC (\ref{e002}) is a penalty on the model complexity. So, intuitively, GIC trades off between the model fitting and  the model complexity by appropriately choosing $a_n$. When $a_n=2$ and $\log n$, (\ref{e002}) becomes the  classical AIC \citep{AIC1973} and BIC \citep{Schwarz_1978},  respectively. The modified BIC \citep{Wangh2009} corresponds to $a_n=C_n \log n$ with a diverging $C_n$ sequence.
The scaled deviation measure and GIC are also studied in \cite{ZLT2010} for regularization parameter selection in a fixed dimensional setting.

%Using GIC,
Our problem of interest now becomes how to appropriately choose $a_n$ such that the tuning parameter $\lambda_0$ can be consistently identified by minimizing (\ref{e002}) with respect to $\lambda$ -- i.e.,  with probability tending to 1 --
\begin{equation}\label{e003}
\inf_{\{\lambda > 0: \alpha_{\lambda} \neq \alpha_0\}}\GIC_{a_n}(\lambda)-\GIC_{a_n}(\lambda_0)>0.
\end{equation}
From  (\ref{e002}) and (\ref{e003}), we can see clearly that to study the choice of $a_n$, it is essential to investigate the asymptotic properties of $D(\hbmu_{\lambda};\bY)$  uniformly over  a range of $\lambda$. Directly studying $D(\hbmu_{\lambda};\bY)$ is challenging because $\hbmu_{\lambda}$ depends on $\hbbeta^{\lambda}$, which is the maximizer of a possibly non-concave function (\ref{e001});  thus, it takes no explicit form,  and more critically, its uniform asymptotic properties are difficult to establish. To overcome these difficulties, we introduce a proxy of $\GIC_{a_n}(\lambda)$,  which is defined as
\begin{align} \label{eq:GIC1}
\GIC^*_{a_n}(\alpha) = \frac{1}{n}\{D(\hbmu^*_\alpha; \bY) + a_n |\alpha|\}
\end{align}
for a given model support $\alpha \subset \{1,\cdots, p\}$ that collects indices of all included covariates, and $\hbmu_\alpha^* = \bb'(\bX\hbbeta^*(\alpha))$ with $\hbbeta^*(\alpha)$ being the unpenalized maximum likelihood estimator (MLE) restricted to the space $\{\bbeta\in\mathbf{R}^p: \supp(\bbeta)=\alpha\}$; that is,
 \begin{equation}\label{eq:mlebeta}
 \hbbeta^*(\alpha) = \arg\max\limits_{\{{\scriptsize\bbeta}\in\mathbf{R}^p:\supp({\scriptsize{\bbeta}})=\alpha\}}\ell_n(\bbeta).
 \end{equation}

The critical difference between (\ref{e002}) and (\ref{eq:GIC1}) is that $\GIC_{a_n}(\lambda)$ is a function of $\lambda$ depending on the  penalized MLE $\hbbeta^{\lambda}$, while $\GIC_{a_n}^*(\alpha)$ is a function of model $\alpha$ depending on the corresponding unpenalized MLE $\hbbeta^*(\alpha)$.  Under some signal-strength assumptions and some regularity conditions, $\hbbeta^{\lambda_0}$ and $\hbbeta^*(\alpha_0)$ are  close to each other asymptotically \citep{zhang06, FanLi_2001_JASA,lv1}. As a consequence, $\GIC_{a_n}(\lambda_0)$ and $\GIC_{a_n}^*(\alpha_0)$ are also asymptotically close, as formally presented in the following proposition:

\begin{proposition}\label{P4} Under Conditions \ref{assp2}, \ref{assp4}, and \ref{assp8} in Section \ref{sec:results}, if $p'_{\lambda_0}(\frac{1}{2}\min_{j\in \alpha_0}|\beta_{0j}|) =o( s^{-1/2}n^{-1/2}a_n^{1/2})$, then
\begin{align}\label{e036}
\GIC_{a_n}(\lambda_0) - \GIC_{a_n}^*(\alpha_0) = o_p(n^{-1}a_n).
\end{align}
\end{proposition}
Furthermore,  it follows from the definition of $\hbbeta^*(\alpha)$ that for any $\lambda>0$,
$
\GIC_{a_n}(\lambda) \geq \GIC_{a_n}^*(\alpha_\lambda).
$
Therefore, Proposition \ref{P4} entails
\begin{align} \nonumber
 \GIC_{a_n}(\lambda) - \GIC_{a_n}(\lambda_0)&\geq \big(\GIC_{a_n}^*(\alpha_\lambda)- \GIC_{a_n}^*(\alpha_0)\big) + \big(\GIC_{a_n}^*(\alpha_0) - \GIC_{a_n}(\lambda_0)\big) \\
 &= \big(\GIC_{a_n}^*(\alpha_\lambda)- \GIC_{a_n}^*(\alpha_0)\big) +o_p(n^{-1}a_n).
 \label{e053}
\end{align}
Hence, the difficulties of directly studying $\GIC$ can be overcome by using the proxy $\GIC^*$ as a bridge, whose properties are elaborated in the next section.

\section{Asymptotic Properties of the Proxy $\GIC$}\label{sec:proxyGIC}
\setcounter{equation}{0}

\subsection{Underfitted Models}\label{sec:underfit}

 From  (\ref{eq:GIC1}), the properties of $\GIC^*$ depend upon the unpenalized MLE $\hbbeta^*(\alpha)$  and scaled deviance measure $D(\hbmu^*_\alpha; \bY)$.
When the truth $\alpha_0$ is given, it is well known from  classical statistical theory that $\hbbeta^*(\alpha_0)$ consistently estimates the population parameter $\bbeta_0$. However, such a result is  less intuitive if $\alpha \neq \alpha_0$. In fact, as shown in Proposition \ref{P1} in Section \ref{sec:results}, uniformly for all $|\alpha|\leq K$ for some positive integer $K > s$ and $K=o(n)$, $\hbbeta^*(\alpha)$ converges in probability to the minimizer $\bbeta^*(\alpha)$ of the following Kullback-Leibler (KL) divergence:
 \begin{align}
I( \bbeta(\alpha))= E\big[\log\big(f^*/g_{\alpha}\big) \big]  =\sum_{i=1}^n \big\{b'(\bx_i\t\bbeta_0)\bx_i^T\big(\bbeta_0-\bbeta(\alpha)\big)-b(\bx_i\t\bbeta_0) + b(\bx_i^T\bbeta(\alpha))\big\},
\label{e087}
\end{align}
where $\bbeta(\alpha)$ is a $p$-dimensional parameter vector with support $\alpha$, $f^*$ is the density of the underlying true model, and $g_{\alpha}$ is the density of the model with population parameter $\bbeta(\alpha)$.
Intuitively,   model $\alpha$  coupled with the population parameter $\bbeta^*(\alpha)$  has the smallest KL divergence from the truth among all models with support $\alpha$. Since the KL divergence is non-negative and $I(\bbeta_0)=0$, the true parameter $\bbeta_0$ is a global minimizer of (\ref{e087}). To ensure identifiability, we assume that (\ref{e087}) has a unique minimizer $\bbeta^*(\alpha)$ for every $\alpha$ satisfying $|\alpha| \leq K$. This unique minimizer assumption will be further discussed in Section \ref{sec:results}. Thus, it follows immediately that $\bbeta^*(\alpha)=\bbeta_0$ for all $\alpha\supseteq\alpha_0$ with $|\alpha| \leq K$, and consequently, $I(\bbeta^*(\alpha))=0$. Hereinafter, we refer to the population model $\alpha$ as the  one associated with the population parameter $\bbeta^*(\alpha)$. We refer to $\alpha$ as an overfitted model if $\alpha \supsetneq\alpha_0$, and as an underfitted model if $\alpha\not\supset\alpha_0$.

For an underfitted population model $\alpha$, the KL divergence $I(\bbeta^*(\alpha)) $ measures the deviance  from the truth due to missing at least one true covariate.  Therefore, we define
\begin{align}\label{e010}
\delta_n = \inf_{\alpha \not\supset \alpha_0 \atop |\alpha|\leq K} \frac{1}{n}  I(\bbeta^*(\alpha))
\end{align}
 as an essential measure of the smallest signal strength of the true covariates, which effectively controls the extent to which the true model can be distinguished from underfitted models.

Let $\bmu_{\alpha}^* = \bb'(\bX\bbeta^*(\alpha))$ and $\bmu_0 = \bb'(\bX\bbeta_0)$ be the population mean vectors corresponding to the parameter $\bbeta^*(\alpha)$ and the true parameter $\bbeta_0$,  respectively.  It can be seen from definition (\ref{e087}) that
\[
I(\bbeta^*(\alpha))= \frac{1}{2}E[D(\bmu_{\alpha}^*;\bY) - D(\bmu_0;\bY)]
.\]
 Hence, $2I(\bbeta^*(\alpha))$ is the population version of the difference between   $D(\hbmu_{\alpha}^*; \bY)$ and  $D(\hbmu_0^*; \bY)$, where $\hbmu_0^* = \hbmu_{\alpha_0}^*= \bb'(\bX\hbbeta^*(\alpha_0))$ is the estimated population mean vector knowing the truth $\alpha_0$.
 Therefore, the KL divergence $I(\cdot)$ can be intuitively understood as a population distance between a model  $\alpha$ and the truth $\alpha_0$.
 The following theorem formally characterizes the uniform convergence result of the difference of scaled deviance measures to its population version $2I(\bbeta^*(\alpha))$.

\begin{theorem}\label{C1}  Under Conditions \ref{assp2}
%, \ref{assp3},
and \ref{assp4} in Section \ref{sec:results}, as $n\rightarrow \infty$,
\[
\sup_{|\alpha|\leq K  \atop \alpha \subset \{1,\cdots, p\}}\frac{1}{n|\alpha|}\big|D(\hbmu_{\alpha}^*; \bY) - D(\hbmu_0^*; \bY)-2I(\bbeta^*(\alpha))\big| = O_p(R_n),
\]
when either  a) the   $Y_i$'s are bounded or Gaussian distributed, $R_n = \sqrt{(\log p)/n}$, and $\log p = o(n)$; or b) the $Y_i$'s are unbounded and non-Gaussian distributed, the design matrix satisfies $\max_{ij}|x_{ij}| = O(n^{\frac{1}{2}-\tau})$ with $\tau \in (0,1/2]$, Condition \ref{assp5} holds,  $R_n = \sqrt{(\log p)/n} + m_n^2(\log p)/n$,  and $\log p=o(\min\{n^{2\tau}(\log n)^{-1}K^{-2}, n m_n^{-2}\})$ with $m_n$ defined in Condition \ref{assp5}.
\end{theorem}
%
%By its definition in (\ref{e023}), a smaller value of $D(\hbmu_{\alpha}^*; \bY)$ indicates a better fit of the estimated model to the data.
 Theorem \ref{C1} ensures that for any model $\alpha$ satisfying $|\alpha| \leq K$,
\begin{align}\label{e024}
&\GIC_{a_n}^*(\alpha) - \GIC_{a_n}^*(\alpha_0)= \frac{2}{n}I(\bbeta^*(\alpha))+ (|\alpha| - |\alpha_0|)\big(a_nn^{-1} - O_p(R_n) \big).
\end{align}
 Hence,  it implies that if a model $\alpha$ is far away from the truth -- i.e., $I(\bbeta^*(\alpha))$ is large -- then this population discrepancy can be detected by looking at the sample value of the proxy $\GIC_{a_n}^*(\alpha)$.

Combining (\ref{e010}) with (\ref{e024}), we immediately find that
if $\delta_nK^{-1}R_n^{-1} \rightarrow\infty$ as $n\rightarrow \infty$ and $a_n$ is chosen such that $a_n = o(s^{-1}n\delta_n)$,  then, for large enough $n$,
\begin{align}\label{e013}
\inf_{\alpha\not\supset \alpha_0, |\alpha|\leq K}\GIC_{a_n}^*(\alpha) - \GIC_{a_n}^*(\alpha_0) > \delta_n - sa_nn^{-1} - O_p(KR_n)\geq \delta_n/2,
\end{align}
with probability tending to 1.
Thus, (\ref{e013}) indicates that as long as the signal $\delta_n$ is not decaying to zero too fast,  any underfitted model leads to a non-negligible increment in the proxy $\GIC^*$.  This guarantees that %the true model $\alpha_0$ is not dominated by any underfitted model, so that
minimizing $\GIC_{a_n}^*(\alpha)$ with respect to $\alpha$ can identify the true model $\alpha_0$ among all underfitted models asymptotically.

On the other hand, however,  for any overfitted model $\alpha\supsetneq \alpha_0$ with $|\alpha|\leq K$, $\bbeta^*(\alpha)=\bbeta_0$,  and thus $I(\bbeta^*(\alpha)) = 0$. Consequently, the true model $\alpha_0$ cannot be differentiated from an overfitted model $\alpha$ using the formulation (\ref{e024}). In fact, the study of overfitted models is far more difficult in a high-dimensional setting, as detailed in the next subsection.

\subsection{Overfitted Models: The Main Challenge} \label{sec:overfitted}
%We now characterize overfitted models.
It is known that for an overfitted model $\alpha$, the difference of scaled deviation measures
\begin{equation}\label{e011}
D(\hbmu^*_\alpha; \bY)-D(\hbmu_0; \bY)=2(\ell_n(\hbmu_0^*;\bY) - \ell_n(\hbmu_{\alpha}^*; \bY))
\end{equation}
follows asymptotically the  $\chi^2$ distribution with $|\alpha|-|\alpha_0|$ degrees of freedom when $p$ is fixed. Since there are only a finite number of candidate models for fixed $p$, a model complexity penalty  diverging to infinity at an appropriate rate with sample size $n$ facilitates an information criterion to identify the true model consistently;  see, for example, \cite{Shao_1997_Sinica}, \cite{Baietal_1999_JSPI}, \cite{Wangetal_2007_Bioka}, \cite{ZLT2010}, and references therein.
However, when $p$ grows with $n$, the device in traditional model selection theory cannot be carried forward. Substantial challenges arise from two aspects. One is how to characterize the asymptotic probabilistic behavior of  (\ref{e011}) when  $|\alpha|-|\alpha_0|$ itself is diverging.  The other is how to deal with so many candidate models, the number of which grows combinatorially fast with  $p$.

Let $\bH_0 = \text{diag}\{\bb''(\bX\bbeta_0)\}$ be the diagonal matrix of the variance of $\bY$, and $\bX_{\alpha}$ be a submatrix of $\bX$ formed by columns whose indices are in $\alpha$. For any overfitted model $\alpha$, we define the associated projection matrix as
\begin{equation}\label{e055}
\bB_\alpha=\bH_0^{1/2}\bX_\alpha(\bX_\alpha^T\bH_0\bX_\alpha)^{-1}\bX_\alpha^T\bH_0^{1/2}.
\end{equation}
When the $Y_i$'s are Gaussian,  $\hbbeta^*(\alpha)$ is the least-squares estimate and admits an explicit form so that % It follows from
direct calculations yield
\begin{equation}\label{e018}
D(\hbmu^*_\alpha; \bY)-D(\hbmu_0; \bY) = -(\bY-\bmu_0)^T\bH_0^{-1/2}\big( \bB_{\alpha} - \bB_{\alpha_0} \big)\bH_0^{-1/2}(\bY-\bmu_0).
\end{equation}
When the $Y_i$'s are non-Gaussian, the above result still holds,  but only approximately. In fact, as formally shown in Proposition \ref{P3} in Section 6,
\begin{align} \nonumber
D(\hbmu^*_\alpha; \bY)-D(\hbmu_0; \bY) = & -(\bY-\bmu_0)^T\bH_0^{-1/2} \Big(\bB_{\alpha} - \bB_{\alpha_0}\Big) \bH_0^{-1/2}(\bY-\bmu_0) \\
&+ (|\alpha|-|\alpha_0|) (\text{uniformly small term}). \label{eq:interim}
\end{align}
%As detailed in the Appendix,
The interim result (\ref{eq:interim}) facilitates characterizing the deviation result for the scaled deviance measures by concentrating  on the asymptotic property of \[Z_\alpha=(\bY-\bmu_0)^T\bH_0^{-1/2} (\bB_{\alpha} - \bB_{\alpha_0}) \bH_0^{-1/2}(\bY-\bmu_0).\] When the $Y_i$'s are Gaussian, it can be seen that $Z_{\alpha} \sim \chi^2_{|\alpha|-|\alpha_0|}$ for each fixed $\alpha$. Thus, the deviation result on $\max_{\alpha \supset\alpha_0,|\alpha|\leq K}Z_{\alpha}$ can be obtained by explicitly calculating the tail probabilities of $\chi^2$ random variables. However, if $Y_i$'s are non-Gaussian, it is challenging to study the asymptotic property of $Z_{\alpha}$, not to mention the uniform result across all overfitted models.
To overcome this difficulty, we use the decoupling inequality \citep{Pena1994} to study $Z_{\alpha}$.
The main results for overfitted models are given in the following theorem.

\begin{theorem}\label{T2} Suppose that the design matrix satisfies $\max_{ij}|x_{ij}| = O(n^{\frac{1}{2}-\tau})$ with $\tau \in (1/3,1/2]$ and $\log p =O(n^\kappa)$ for some $0<\kappa<1$. Under Conditions \ref{assp2}--\ref{assp4} in Section \ref{sec:results}, as $n\to\infty$,
 \[
    \frac{1}{|\alpha|-|\alpha_0|}\Big(D(\hbmu^*_\alpha; \bY)-D(\hbmu_0; \bY)\Big)=O_p(\psi_n)
    \]
    uniformly for all $\alpha\supsetneq \alpha_0$ with $|\alpha|\leq K$, and $\psi_n$ is specified respectively in the following two situations:

\begin{itemize}
\item[a)] $\psi_n=\sqrt{\log p}$ when the $Y_i$'s are bounded, $K  = O ( \min\{n^{(3\tau-\kappa-1)/3}, n^{(4\tau -1- 3\kappa)/8}\})$ and  $\kappa \leq 3\tau-1$;

\item[b)]  $\psi_n=\log p$ when the $Y_i$'s are Gaussian distributed; or when the $Y_i$'s are unbounded non-Gaussian distributed, additional Condition \ref{assp5}  holds, $K=O \big( n^{ (6\tau-2-\kappa)/6}  (\sqrt{\log n} + m_n)^{-1}\big)$, $\kappa \leq 6\tau -2$, and $m_n=o\big( n^{(6\tau-2-\kappa)/6}\big)$.
\end{itemize}
\end{theorem}
%
%Theorem \ref{T2} clearly illustrates  the impact of the distribution of the response variables on the scaled deviance measures.
The results in Theorem \ref{T2} hold for all overfitted models, which provides an insight into a high-dimensional scenario beyond the asymptotic  result characterized by $\chi^2$ distribution when $p$ is fixed. Theorem \ref{T2} entails that when $a_n$ is chosen  such that $a_n\psi_n^{-1} \rightarrow \infty$, uniformly for any overfitted model $\alpha \supsetneq \alpha_0$,
\begin{align}\label{e021}
\GIC_{a_n}^*(\alpha) - \GIC_{a_n}^*(\alpha_0) = \frac{|\alpha|-|\alpha_0|}{n}\left
\{a_n-O_p(\psi_n)\right\} > a_n/(2n)
\end{align}
 with asymptotic probability 1.
Thus, we are now able to  differentiate overfitted models from the truth  by examining  the values of the proxy $\GIC_{a_n}^*(\alpha) $.

%\subsection{Consistent Model Selection with Proxy GIC}\label{sec:GIC}
%
%
%\setcounter{equation}{0}
%
%Combining the results presented in the previous two subsections we have the following theorem.
%
%\begin{theorem}\label{T3}
% Under the same Conditions in Theorem \ref{C1} and Theorem {\ref{T2}}, as $n\to \infty$
%  \begin{align*}
%P\left(\inf_{\alpha\supsetneq\alpha_0}\GIC^*_{a_n}(\alpha)-\GIC^*_{a_n}(\alpha_0) >\frac{\delta_n}{2} \text{ and } \inf_{\alpha\not\supset\alpha_0}\GIC^*_{a_n}(\alpha)-\GIC^*_{a_n}(\alpha_0)>\frac{a_n}{2n} \right) \longrightarrow 1,
%\end{align*} if $\delta_nK^{-1} R_n^{-1} \rightarrow \infty$,  $a_n$ satisfies
%$n \delta_n s^{-1} a_n^{-1}\to \infty$ and $a_n \psi_n^{-1}\to \infty$, where $R_n$ and $\psi_n$ are specified in Theorems \ref{C1} and \ref{T2}.
%\end{theorem}

\section{Consistent Tuning Parameter Selection with GIC}\label{sec:tune}

\setcounter{equation}{0}

%
% Let $\lambda_{\max}$ and $\lambda_{\min}$ be respectively the upper and lower limits of the regularization parameter chosen in the way such that $\lambda_0\in [\lambda_{\min}, \lambda_{\max}]$, $\alpha_{\lambda_{\max}}$ is empty and  $\hbbeta^{\lambda_{\min}}$ is sparse with the corresponding model size $|\alpha_{\lambda_{\min}}|=o(n)$. As discussed in the introduction, most research on tuning parameter selection has been restrained to the $p\leq n$ or linear regression setting. Our main objective is on how to identify $\lambda_0 \in [\lambda_{\min}, \lambda_{\max}]$ such that the true model is consistently recovered in the ultra-high dimensional generalized linear model setting.

Now, we are ready to study the appropriate choice of $a_n$ such that the tuning parameter $\lambda_0$ can be selected consistently by minimizing GIC as defined in (\ref{e002}). In practical implementation, the tuning parameter $\lambda$ is considered over a range and,  correspondingly, a collection of models are produced. Let $\lambda_{\max}$ and $\lambda_{\min}$ be, respectively,  the upper and lower limits of the regularization parameter, where $\lambda_{\max}$ can be easily chosen such that $\alpha_{\lambda_{\max}}$ is empty and $\lambda_{\min}$ can be chosen such that
$\hbbeta^{\lambda_{\min}}$ is sparse, and the corresponding model size $K=|\alpha_{\lambda_{\min}}| $ satisfies conditions in Theorem \ref{T4} below.
Using the same notations as those in \cite{ZLT2010}, we partition the interval $[\lambda_{\min}, \lambda_{\max}]$ into subsets
\begin{align*}
&\Omega_- = \{\lambda \in [\lambda_{\min}, \lambda_{\max}]: \alpha_\lambda  \not\supset \alpha_0\}, \\
%& \Omega_0 = \{\lambda \in [\lambda_{\min}, \lambda_{\max}]: \alpha_\lambda = \alpha_0 \}, \text{ and }\\
& \Omega_+ = \{\lambda \in [\lambda_{\min}, \lambda_{\max}]: \alpha_\lambda \supset \alpha_0 \text{ and } \alpha_\lambda \neq \alpha_0\}.
\end{align*}
Thus, $\Omega_-$ is the set of $\lambda$'s that result in underfitted models, and $\Omega_+$ is the set of $\lambda$'s that produce overfitted models.

We now present the main result of the paper. Combining (\ref{e053}), (\ref{e013}), and (\ref{e021}) with Proposition \ref{P4}, we have the following theorem.

\begin{theorem}\label{T4}   Under the same conditions in Proposition \ref{P4}, Theorem \ref{C1}, and Theorem {\ref{T2}}, if $\delta_nK^{-1} R_n^{-1} \rightarrow \infty$,  $a_n$ satisfies
$n \delta_n s^{-1} a_n^{-1}\to \infty$ and $a_n \psi_n^{-1}\to \infty$,  where $R_n$ and $\psi_n$ are specified in Theorems \ref{C1} and \ref{T2}, then as $n\to \infty$,
\begin{align*}
P\left( \inf_{\lambda \in \Omega_-\cup \Omega_+}\GIC_{a_n}(\lambda) > \GIC_{a_n}(\lambda_0)\right) \rightarrow 1,
\end{align*}
where $\lambda_0$ is the tuning parameter in Condition \ref{assp8}  that consistently identifies the true model. %d.
\end{theorem}

The two requirements on $a_n$ specify a range such that GIC is consistent in model selection for penalized MLEs. They reveal the synthetic impacts due to the signal strength, tail probability behavior of the response, and the dimensionality. Specifically, $a_n \psi_n^{-1}\to \infty$ means that $a_n$ should diverge to $\infty$ adequately fast so that the true model is not dominated by overfitted models.  On the other hand, $n \delta_n s^{-1} a_n^{-1}\to \infty$ restricts the diverging rate of $a_n$, which can be viewed as constraints due to the signal strength quantified by $\delta_n$ in (\ref{e010}) and the size $s$ of the true model.

Note that $\psi_n$ in Theorem \ref{T2} is a power of $\log p$. The condition $a_n\psi_n^{-1}$ in Theorem \ref{T4} clearly demonstrates the impact of dimensionality $p$ so that the penalty on the model complexity should incorporate $\log p$.  From this perspective, the AIC and even the BIC may fail to consistently identify the true model when $p$ grows exponentially fast with $n$.  As can be seen from the technical proofs in the Appendix,  the huge number of overfitted candidate models is the  leads to the model complexity penalty involving $\log p$. Moreover,  Theorem \ref{T4} actually accommodates the existing results -- for example, the modified BIC as in \cite{Wangh2009}.
If dimensionality $p$ is only of polynomial order of sample size $n$ ( i.e., $p = n^{c}$ for some $c \ge 0$), then $\log p = O(\log n)$, and thus the modified BIC with $a_n = (\log \log n) \log n$ can consistently select the true model in Gaussian linear models. As mentioned in the introduction, Theorem \ref{T4}
produces  a phase diagram of how the model complexity penalty should adapt to the growth of sample size $n$ and dimensionality $p$.

Theorem \ref{T4} specifies a  range of $a_n$ for consistent model selection:
\[
n \delta_n s^{-1} a_n^{-1}\to \infty \text{ and } a_n \psi_n^{-1}\to \infty.
\]
 For practical implementation, we propose to use a uniform choice $a_n=(\log \log n)\log p$ in the GIC.  The diverging part $\log \log n$ ensures $a_n \psi_n^{-1}\to \infty$ for all situations in Theorem \ref{T4}, and the slow diverging rate can ideally avoid underfitting.  As a direct consequence of Theorem \ref{T4}, we have the following corollary for the validity of the choice of $a_n$.

\begin{corollary}\label{C2}
 Under the same Conditions in Theorem  {\ref{T4}} and letting $a_n=(\log \log n)\log p$, as $n\to \infty$
 \begin{align*}
P\left(\inf_{\lambda \in \Omega_-\cup \Omega_+}\GIC_{a_n}(\lambda)>\GIC_{a_n}(\lambda_0)\right) \longrightarrow 1
\end{align*} if $\delta_nK^{-1} R_n^{-1} \to \infty$ and
$n \delta_n s^{-1} (\log \log n)^{-1}(\log p)^{-1}\to \infty$, where $R_n$ is as specified in Theorem \ref{C1}.
\end{corollary}

When $a_n$ is chosen appropriately as in Theorem \ref{T4} and Corollary \ref{C2},
minimizing (\ref{e002})  identifies the tuning parameter $\lambda_0$ with probability tending to one.  This concludes a valid tuning parameter selection approach for identifying the true model for penalized likelihood methods.

\section{Numerical Examples}\label{sec:numeric}
\setcounter{equation}{0}

\subsection{Simulations}
 We implement the proposed tuning parameter selection procedure using GIC with $a_n=(\log \log n)\log p$ as proposed in Corollary \ref{C2}. %and $\gamma_n=\log\log n$.  %Throughout the numerical examples, the $\gamma_n$ is chosen as $\log\log n$.
 We compare its performance with those obtained by using AIC ($a_n=2$) and BIC ($a_n=\log n$).
 In addition $a_n=\log p$ is also assessed, which is one of the possible criteria proposed in \cite{WangZhu_2011_JMVA}.
  Throughout the simulations, the number of replications is 1,000.  In the numerical studies, the performance of AIC is substantially worse than other tuning parameter selection methods, especially when $p$ is much larger than $n$, so that we omit the corresponding results for the ease of presentation.

We first consider the Gaussian linear regression where continuous response variables are generated from the model
\begin{equation}\label{e501}
Y_i=\bx_i^T \bbeta+\epsilon_i, \quad i=1,\dots, n.
\end{equation}
The row vectors $\bx_i$'s of the design matrix $\bX$ are generated independently from a $p$-dimensional %uncorrelated  normal distribution with unit variance
multivariate standard Gaussian distribution, and the $\epsilon_i$'s  are iid $N(0,\sigma^2)$  with $\sigma=3.0$ corresponding to the noise level.  In our simulations, $p$ is taken to be the integer part of $\exp\{(n-20)^{0.37}\}$. We let $n$ increase from $100$ to $500$ with $p$ ranging from $157$ to $18,376$.  The number of true covariates $s$ is growing with $n$ in the following manner.  Initially, $s=3$, and the first 5 elements of the true coefficient vector $\bbeta_0$ are set to be $(3.0,1.5,0.0,0.0,2.0)^T$ and all remaining elements are zero. Afterward, $s$ increases by $1$ for every $40$-unit increment in $n$ and the new element takes the value $2.5$. For each simulated data set, we calculate the penalized MLE $\hbbeta^{\lambda}$ using (\ref{e001}) with $\ell_n(\bbeta)$ being the log-likelihood function for the linear regression model (\ref{e501}).

 We then consider  the logistic regression  where binary response variables are generated from the model
 \begin{equation}\label{e502}
 P(Y_i=1|\bx_i)=1/(1+e^{-\bx_i^T\sbbeta}), \quad i=1,\dots, n.
 \end{equation}
The design matrix $\bX = (\bx_1, \bx_2, \cdots, \bx_n)\t$, dimensionality $p$ and sample size $n$ are samely specified as in the linear regression example.
 The first 5 elements of $\bbeta_0$ are set to be $(-3.0,1.5,0.0,0.0,-2.0)^T$,  and the remaining components are all zeros. Afterward, the number of nonzero parameters $s$ increases by $1$ for every $80$-unit increment in $n$ with the value being $2.0$ and $-2.0$ alternatively. The penalized MLE $\hbbeta^{\lambda}$  is computed for each simulated data set based on (\ref{e001}) with $\ell_n(\bbeta)$ being the log-likelihood function for the logistic regression model (\ref{e502}).

\begin{figure}[htbp]
\begin{tabular}{cc}
\includegraphics[width=0.48\textwidth,height=0.4\textheight]{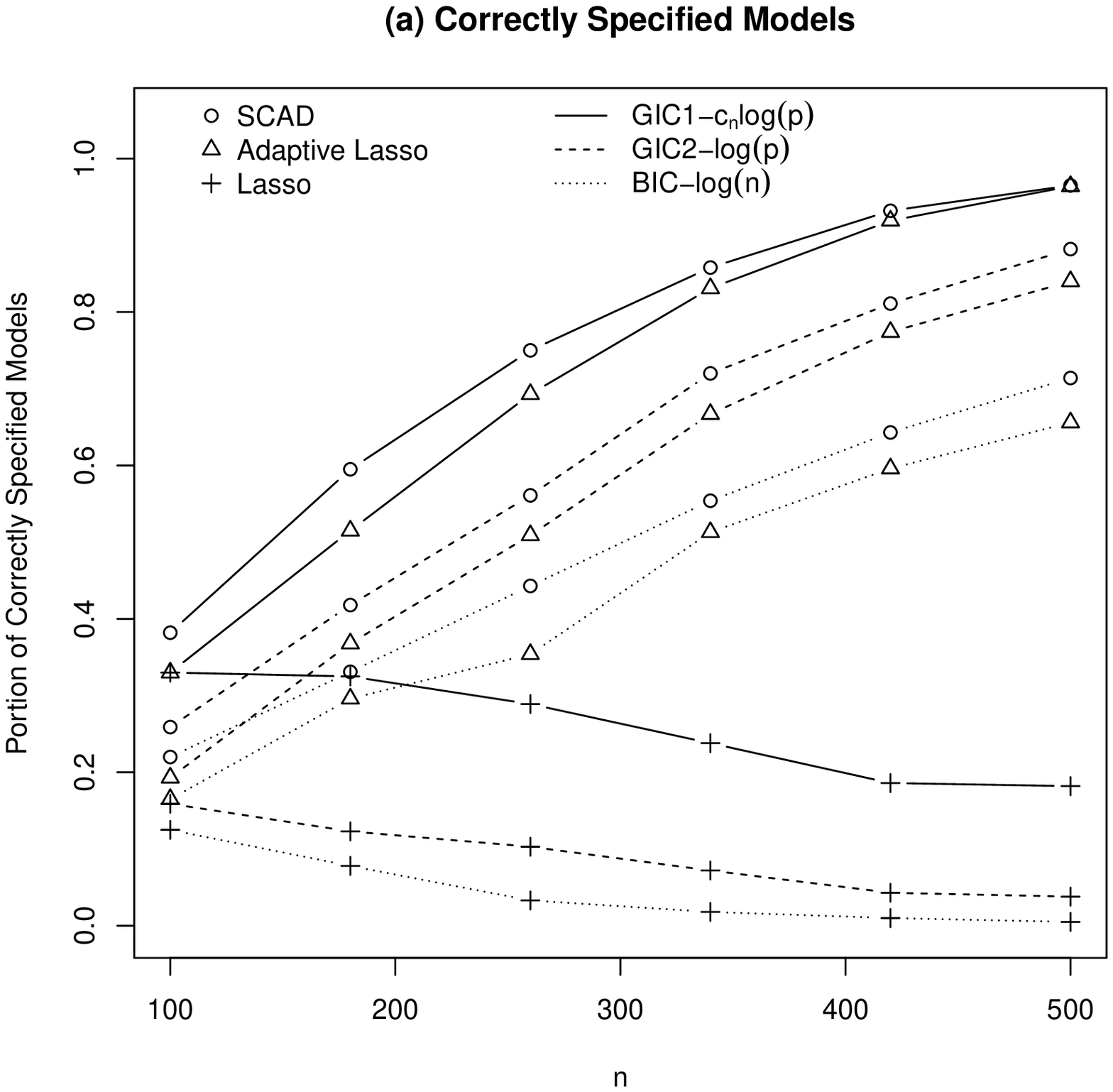}& \includegraphics[width=0.48\textwidth,height=0.4\textheight]{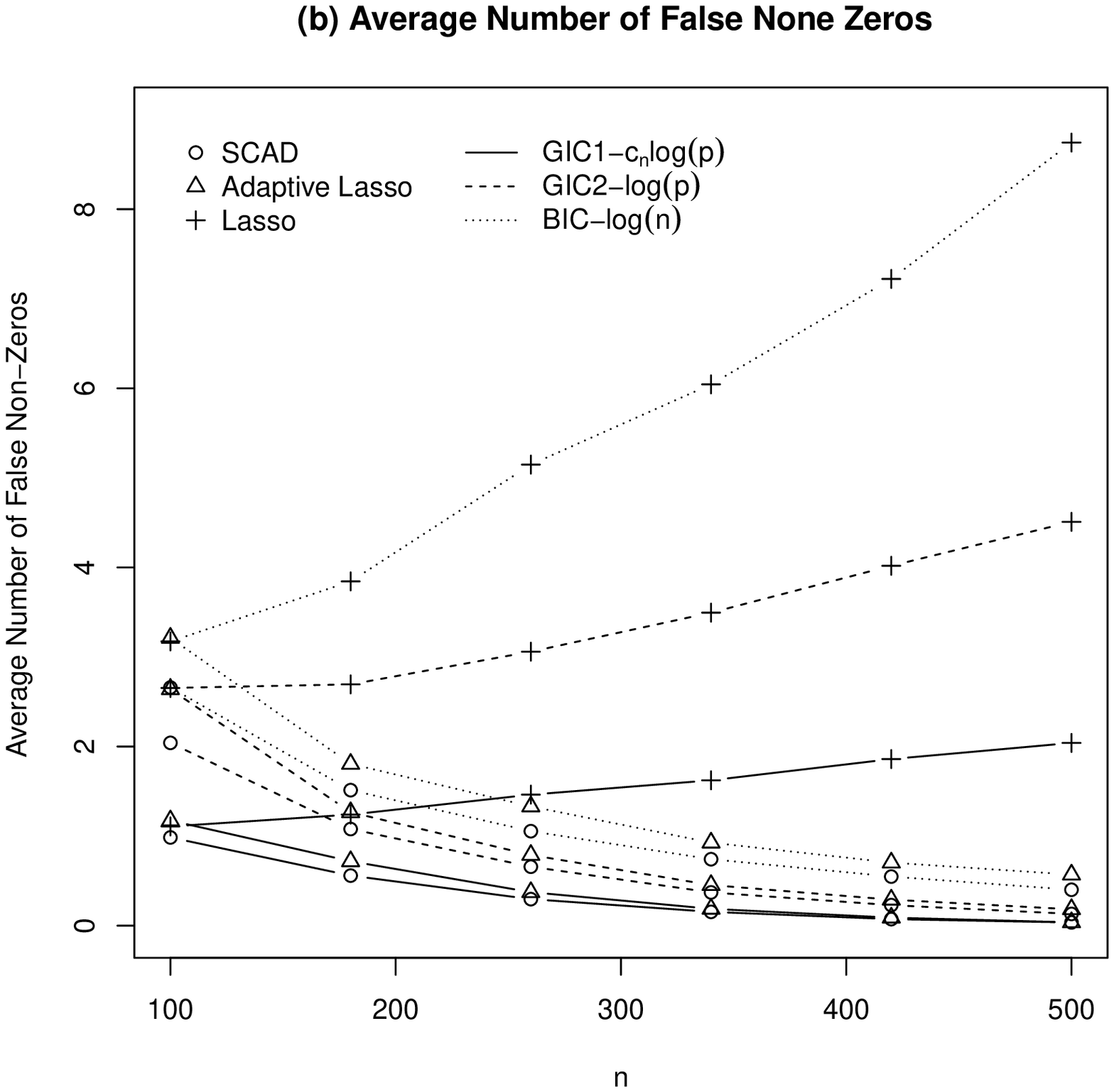}\\
\includegraphics[width=0.48\textwidth,height=0.4\textheight]{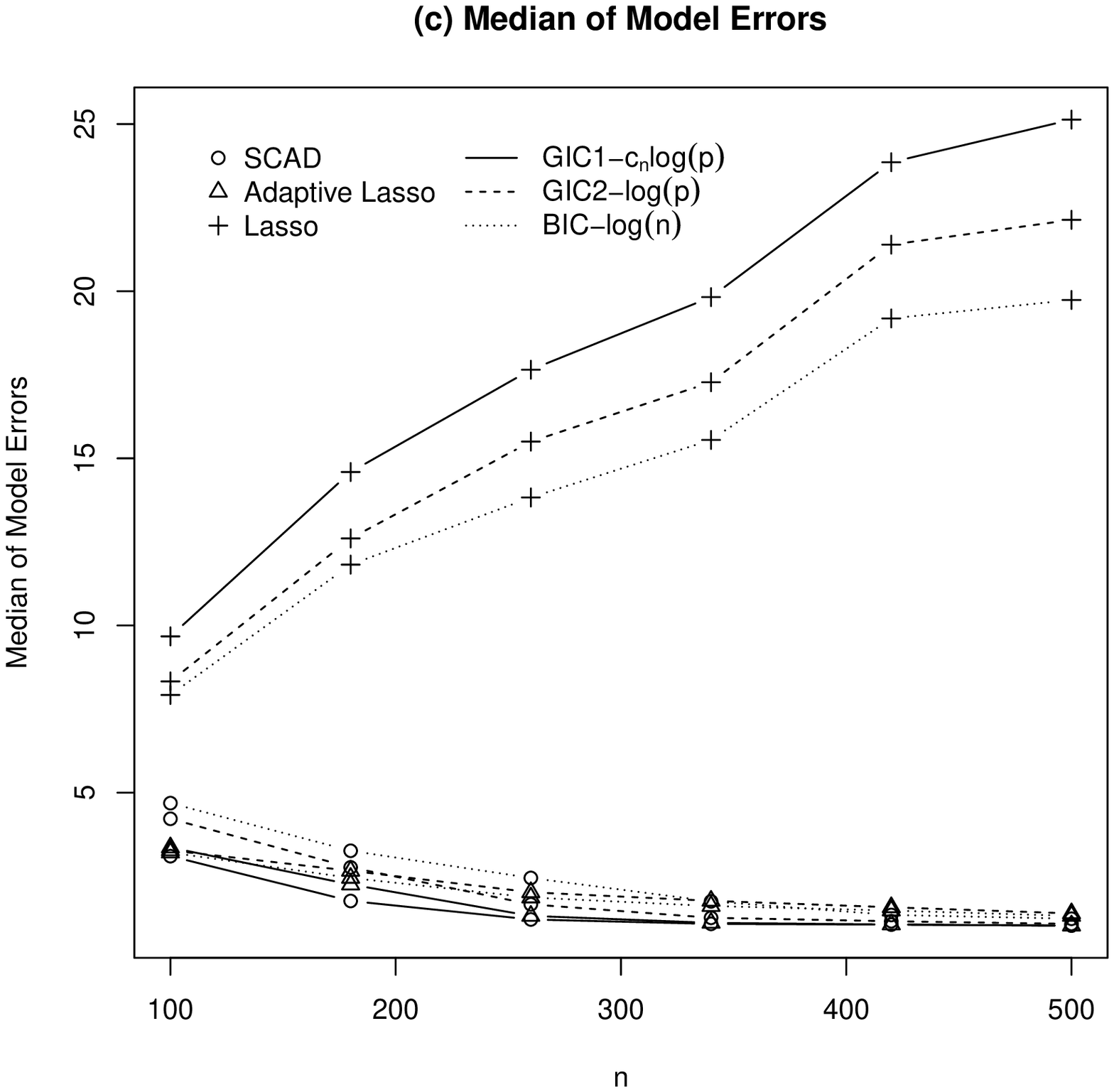}& \includegraphics[width=0.48\textwidth,height=0.4\textheight]{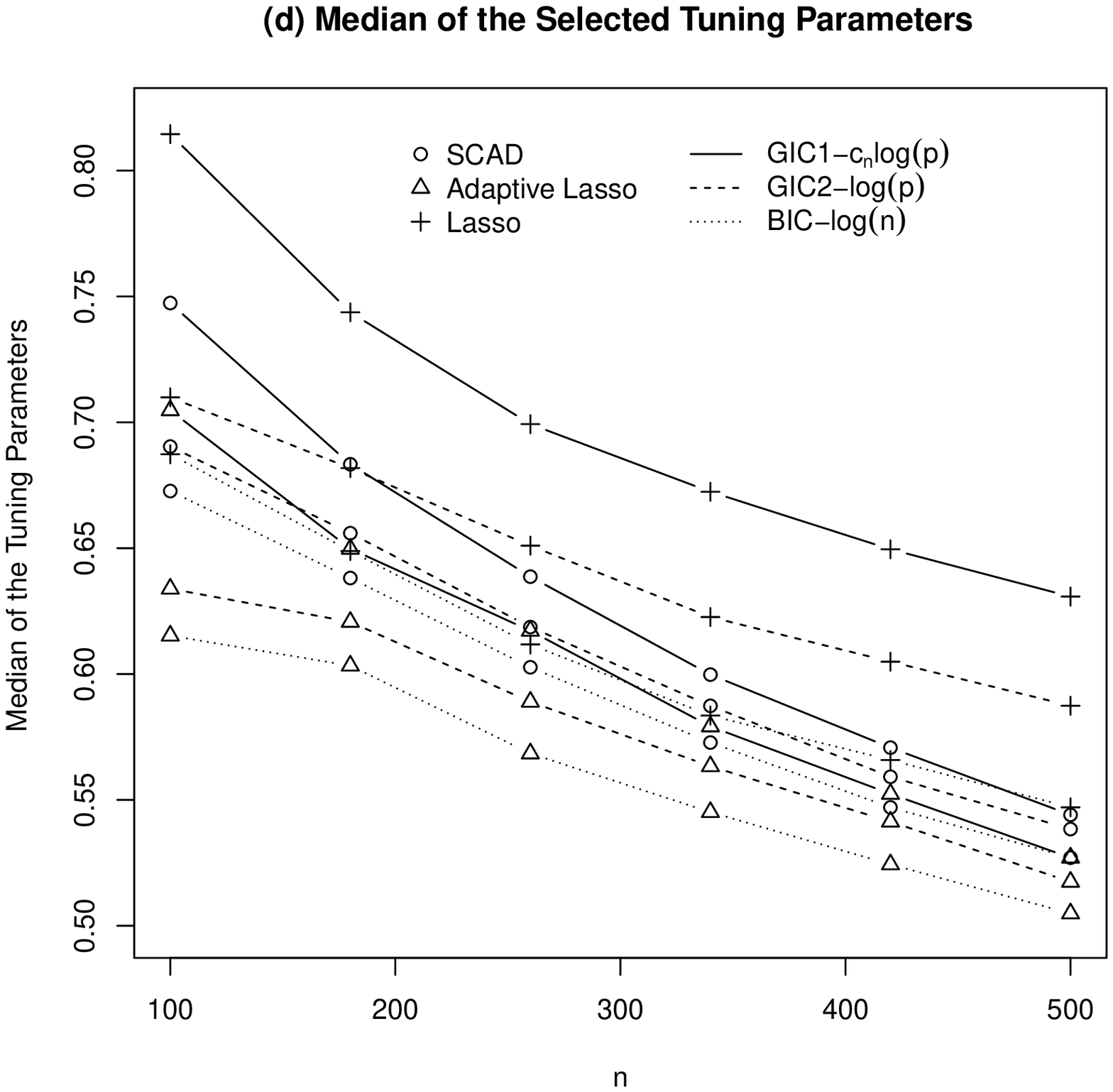}\end{tabular}
\caption{Results for linear model with Gaussian errors and $c_n=\log \log n$. %(a) the percentage of correctly specified models, (b) the average number of false nonzeros identified, (c) the median of relative model error based on estimated unpenalized selected models where ratios of the model errors over the oracle model errors are reported and (d) the median of the selected tuning parameters.
}
    \label{fig1}
\end{figure}

\begin{figure}[htbp]
\begin{tabular}{cc}
\includegraphics[width=0.48\textwidth,height=0.4\textheight]{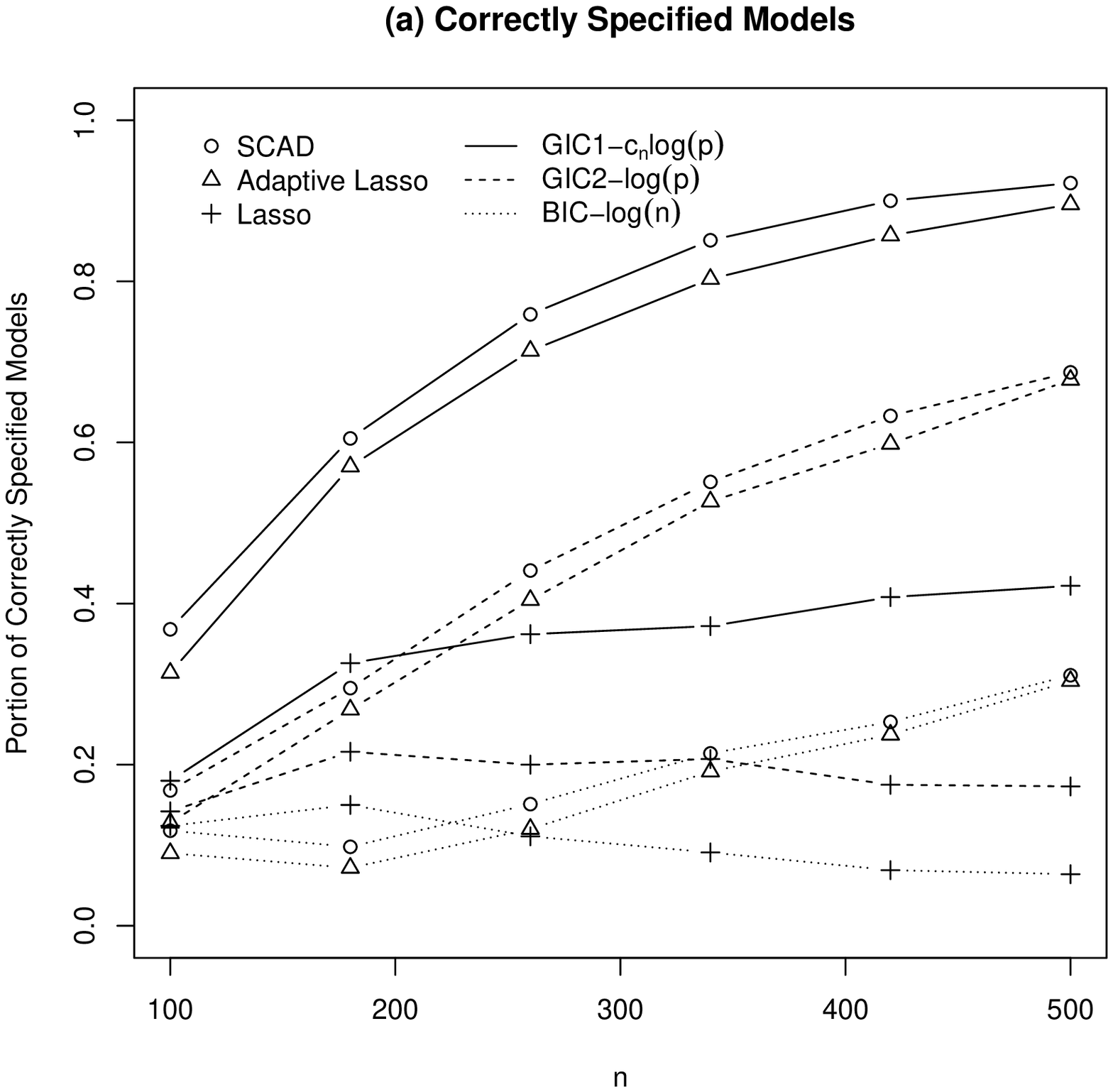}& \includegraphics[width=0.48\textwidth,height=0.4\textheight]{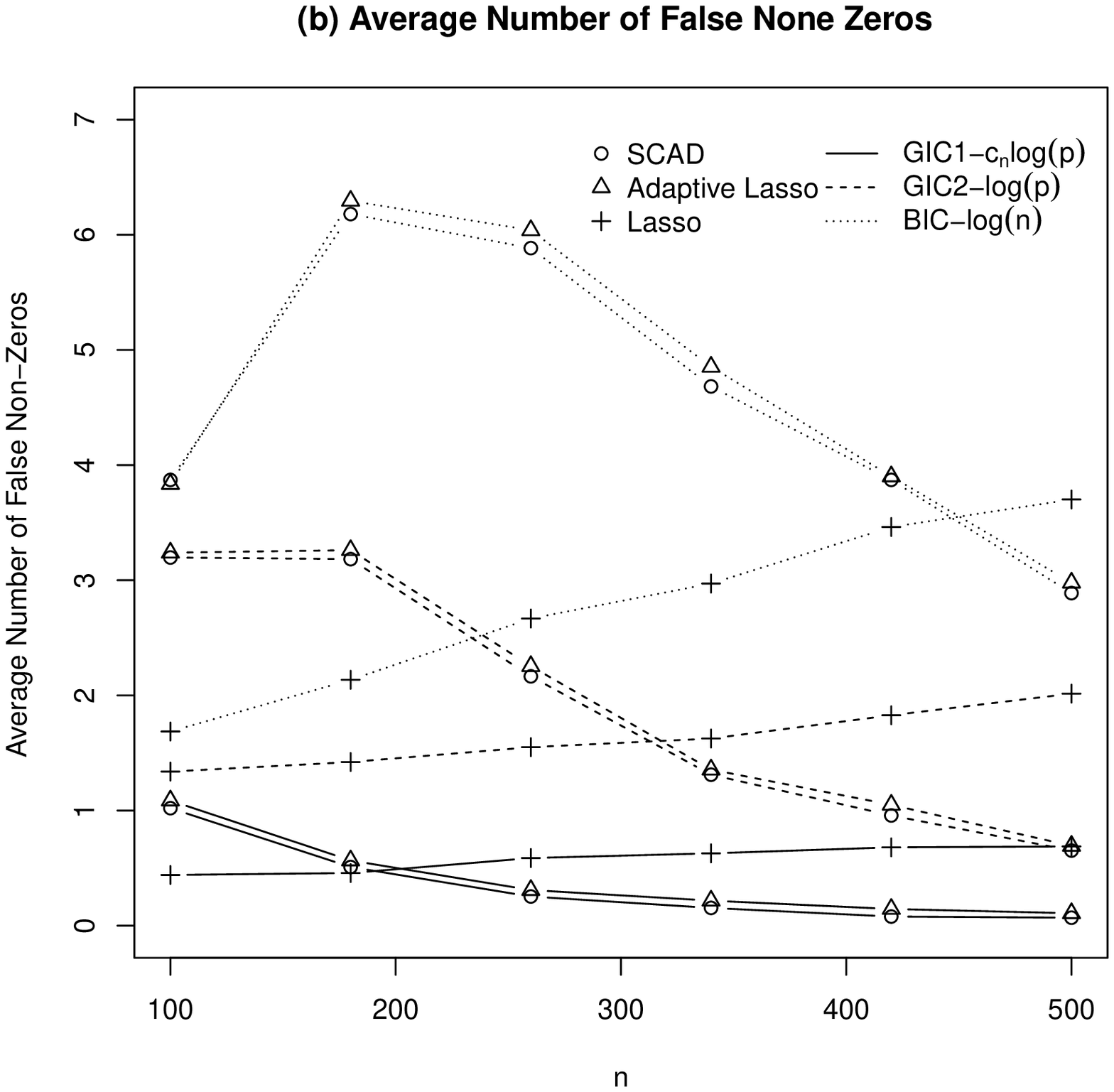}\\
\includegraphics[width=0.48\textwidth,height=0.4\textheight]{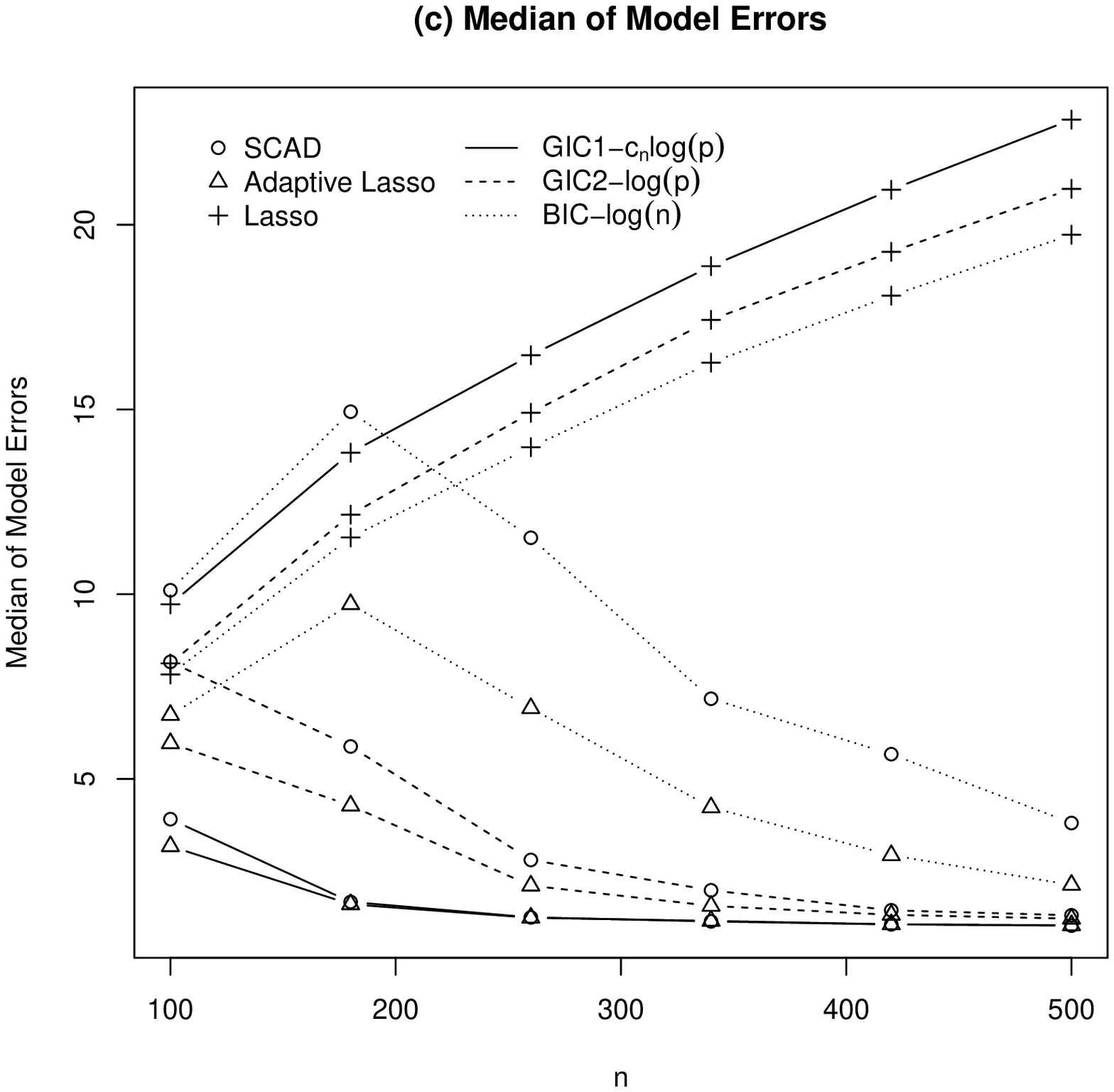}& \includegraphics[width=0.48\textwidth,height=0.4\textheight]{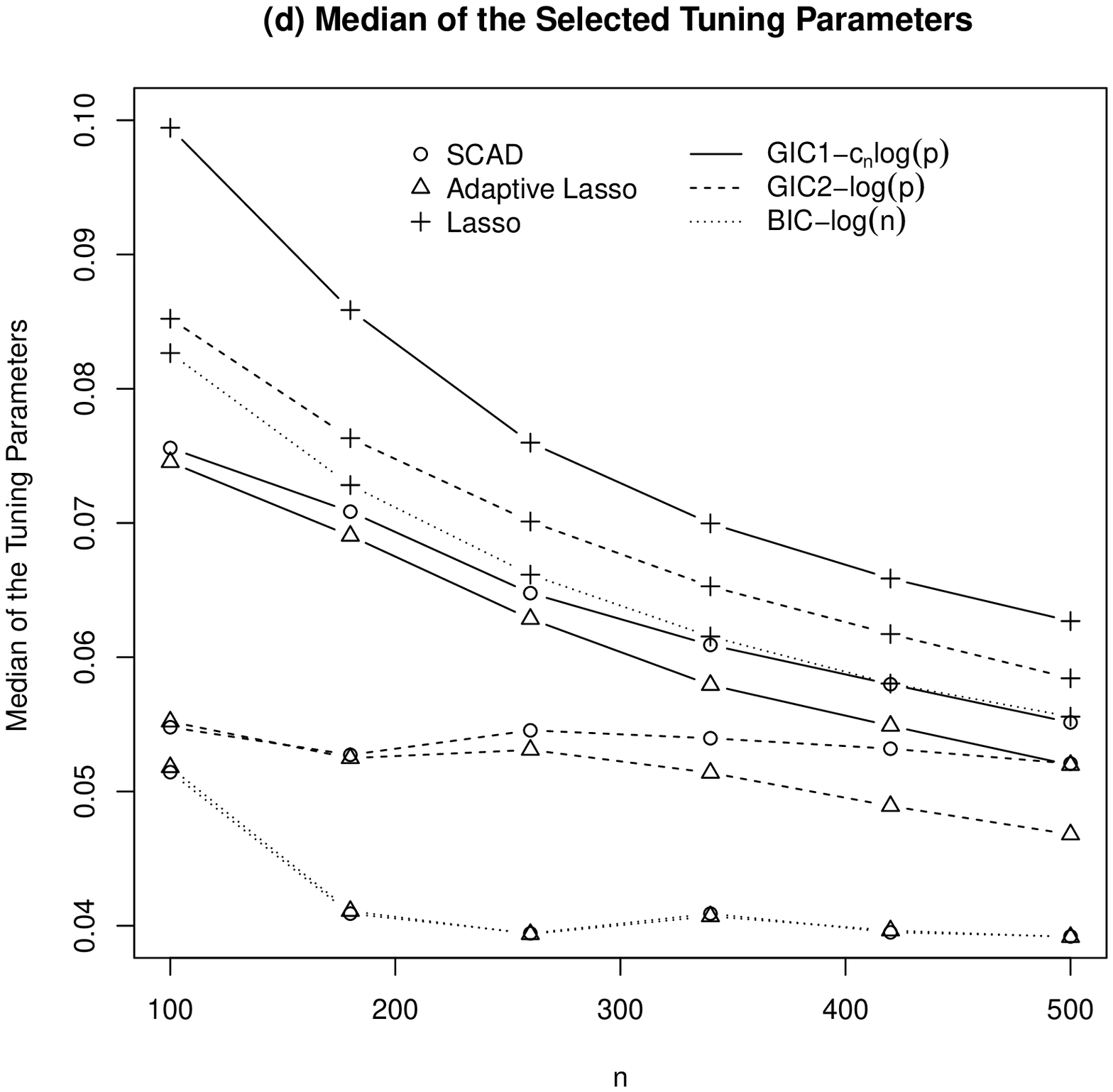}
\end{tabular}
\caption{Results for logistic regressions and $c_n=\log \log n$.  %(a) the percentage of correctly specified models, (b) the average number of false nonzeros identified, (c) the median of relative model error based on estimated unpenalized selected models where ratios of the model errors over the oracle model errors are reported and (d) the median of the selected tuning parameters.
}
    \label{fig2}
\end{figure}

We apply regularization methods with the Lasso \citep{tibshirani2}, SCAD \citep{FanLi_2001_JASA} and MCP \citep{Zhang_2010_AOS} penalties, and the coordinate decent algorithms \citep{BrehenyHuang_2010_AOAS,Friedmanetal_2010_JSS} are carried out in optimizing the objective functions. The results of the MCP penalty are very similar to those of the SCAD penalty, and they are omitted. %to save space.
We also compare with a re-weighted adaptive Lasso method, whose adaptive weight for $\beta_j$ is chosen % by using the Lasso estimator $\hbbeta_{Lasso} = (\hbeta_{1,Lasso},\cdots, \hbeta_{p,Lasso})\t$ as the initial value and choose the weight for adaptive Lasso
as $p'_{\lambda}(\hbeta_{j,Lasso}^\lambda)$ with $p'_{\lambda}(\cdot)$ being the derivative of SCAD penalty, and $\hbbeta_{Lasso}^\lambda = (\hbeta_{1,Lasso}^\lambda,\cdots, \hbeta_{p,Lasso}^\lambda)\t$ being the Lasso estimator. We remark that this re-weighted adaptive Lasso method shares the same spirit as the original SCAD-regularized estimate. In fact, a similar method, the local linear approximation method, has been proposed and studied in \cite{ZL08}. They show that under some conditions of the initial estimator $\hbbeta_{Lasso}$, the re-weighted adaptive Lasso estimator discussed above enjoys the same oracle property as the original SCAD-regularized estimator. The similarities of these two estimates can also be seen in Figures \ref{fig1} and \ref{fig2}. %in our simulations.

For each regularization method -- say, the SCAD method --  when carrying out the tuning parameter selection procedure, we first calculate partly the solution path by choosing $\lambda_{\min}$ and $\lambda_{\max}$. Here, $\lambda_{\max}$ is chosen in such a way that no covariate is selected by the SCAD method in the corresponding model, while $\lambda_{\min}$ is the value where $[3\sqrt{n}]$ covariates are selected.   Subsequently, for a grid of 200 values of $\lambda$ equally spaced on the log scale over $[\lambda_{\min}, \lambda_{\max}]$, we calculate the SCAD-regularized estimates. This results in a sequence of candidate models. Then we apply each of the aforementioned tuning parameter selection methods to select the best model from the sequence. We repeat the same procedure for other regularization methods.

To evaluate the tuning parameter selection methods,
we calculate the percentage of correctly specified models, the average number of false zeros identified, and the median model error $E (\bx_i^T \hbbeta-\bx_i^T\bbeta_0)^2$ for each selected model. We would like to remark that the median and mean of model errors are qualitatively similar, and we use the median just to make results comparable to those in \cite{Wangh2009}. The comparison results are summarized in Figures \ref{fig1} and \ref{fig2}.
We clearly see that for the SCAD and adaptive Lasso, higher percentages of correctly specified models are achieved when $a_n=(\log \log n) \log p$ is used. The Lasso method performs relative poorly, due to its bias issue \citep{FanLv_2009_Sinica}. In fact, our GIC aims at selecting the true model, while it is known that Lasso tends to over-select many variables.  Thus the GIC selects larger values of tuning parameter $\lambda$ for Lasso than for other regularization methods to enforce the model sparsity, as shown in panel (d) of Figures \ref{fig1} and \ref{fig2}. This larger thresholding level $\lambda$ results in an even more severe bias issue as well as missing true weak covariates for Lasso method, which in turn cause larger model errors (see panel (c)).  %We also remark that the bad performance of Lasso with GIC is expected, since the high collinearity in high dimensions makes the irrepresentable condition \citep{ZhaoYu_2006_JMLR} hard to be satisfied, and thus the true model is likely not in the solution path of Lasso. %Since our GIC is constructed to select the true model from a sequence of models containing the true one, in some sense it does not apply to the setting of Lasso method. But for completeness of the story we still include Lasso in the comparisons.

As expected and seen from panel (b),  $a_n=(\log \log n) \log p$ in combination with the SCAD and adaptive Lasso has much smaller false positives, which is the main reason for the substantial improvements in model selection. This demonstrates the need for applying an appropriate value of $a_n$ in ultra-high dimensions.  In panel (c), we report the median of relative model errors of the re-fitted unpenalized estimates for each selected model.   We use the oracle model error from the fitted true model as the baseline, and report the ratios of model errors for selected models to the oracle ones.  From panel (c) of Figures \ref{fig1} and \ref{fig2}, we can see that the median relative model errors  corresponding to $(\log \log n)\log p$  decrease to $1$ very fast, and are consistently smaller  than those using BIC, for both SCAD and adaptive Lasso.
%We  find that the relative model errors of Lasso actually increase indicating a  worsen relative performance, though the absolute values of the errors are decreasing.
This demonstrates the improvement by using  a more accurate model selection procedure in an ultra-high-dimensional setting.    As the sample size $n$ increases, the chosen tuning parameter decreases as shown in panel (d). We also observe from panel (d) that $a_n=(\log \log n) \log p$ results in relatively larger values of selected $\lambda$. Since $\lambda$ controls the sparsity level of the model, panel (d) reflects the extra model complexity penalty made by our GIC in order to select the true model from a huge collection of candidate models, as theoretically demonstrated in previous sections.    %On the other hand, although $a_n=(\log \log n) \log p$ result in even larger tuning parameters for the Lasso, its overall performances in identifying the true model and the relative model errors are substantially worse than those of the MCP and adaptive Lasso. However, the two-step re-weighted adaptive Lasso approach is very promising even the initial estimator can performance poorly.

\subsection{Gene Expression Data Analysis}

\begin{figure}[htbp]
\begin{center}
\includegraphics[width=0.5\textwidth]{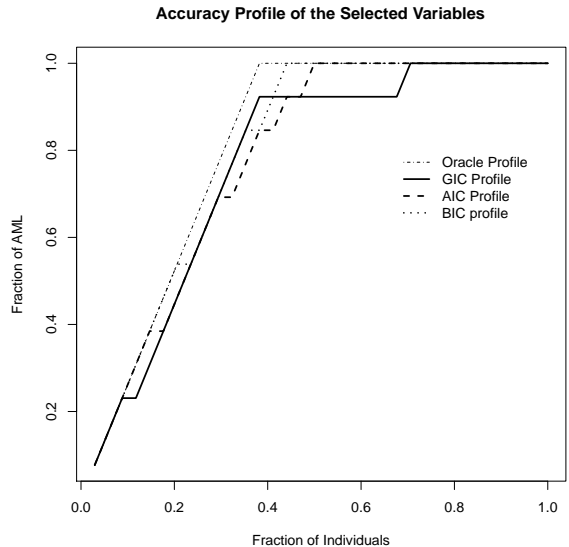}
\end{center}
\caption{Accuracy profiles of the selected gene expression level  for discriminating between the types of leukemias. }
    \label{fig3}
\end{figure}

We then examine the tuning parameter selection procedures on the data from a gene expression study of leukemia patients.  The study  is described in \cite{Golubetal_1999_SCI} and the data-set is available at {\verb+http://www.genome.wi.mit.edu/MPR+}.  The training set contains gene expression levels of two types of acute leukemias: 27 patients with acute lymphoblastic leukemia (ALL) and 11 patients with acute myeloid leukemia (AML).  Gene expression levels for another 34 patients are available in a test set. We applied the pre-processing steps as in \cite{Dudoitetal_2002_JASA}, which resulted in $p=3,051$ genes.
We create a binary response variable based on the types of leukemias by letting $Y_i=1$ (or $0$) if the corresponding patient has ALL (or AML).
  By using the gene expression levels as covariates in $\bx_i$, we  %followed the approach in \cite{BrehenyHuang_2010_AOAS} to
 fit the data to the penalized logistic regression model (\ref{e502})  using the SCAD penalty for a sequence of tuning parameters.  Applying the AIC criterion, $7$ genes were selected, which is close to the results by the cross-validation procedure applied in \cite{BrehenyHuang_2010_AOAS}.
Applying the BIC criterion, $4$ genes were selected.
When applying the GIC criterion with $a_n=(\log\log n) \log p $, only one gene, {\verb+CST3 Cystatin C+} (amyloid angiopathy and cerebral hemorrhage), was selected.
We note that this gene was included in those selected by the AIC and BIC.
Given the small sample size ($n=38$) and extremely high dimensionality ($p>3,000$), the variable selection result is actually not surprising.  By further examining the gene expression level of {\verb+CST3 Cystatin C+}, we can find that it is actually highly informative in differentiating between the two types ALL and AML even using only one gene.
To assess the out-of-sample performance, we generated the accuracy profile by first ordering the patients according to the gene expression level of {\verb+CST3 Cystatin C+} and then plotting the top $x\%$ patients against the $y\%$ of actual AML cases among them.
By looking at the accuracy profile in Figure \ref{fig3} according to the ranking using the gene expression level of {\verb+CST3 Cystatin C+}, we can see that the profile is very close to the oracle profile that knows the truth.
For comparisons, we also plot the accuracy profiles based on genes selected by the AIC and BIC for comparisons.
As remarked in \cite{Dudoitetal_2002_JASA}, the out-of-sample test set is more heterogeneous because of a broader range of samples, including those from peripheral blood  and bone marrow, from childhood AML patients, and even from laboratories that used different
sample preparation protocols. In this case,  the accuracy profile  %than the out-of-sample error rate
is instead an informative indication in telling the predicting power of gene expression levels.

\section{Technical Conditions and Intermediate Results} \label{sec:results}
\setcounter{equation}{0}

%Recall that $\bbeta^*(\alpha)$ is the unique minimizer of KL divergence (\ref{e087}).
 For the model identifiability, we assume in Section \ref{sec:underfit} that (\ref{e087}) has a unique minimizer $\hbbeta^*(\alpha)$ for all $\alpha$ satisfying $|\alpha|\leq K$.
By optimization theory, $\bbeta^*(\alpha)$ is the unique solution to the first-order equation
\begin{align}\label{e006}
\bX^T_\alpha\big\{\bb'(\bX\bbeta_0)-\bb'\big(\bX\bbeta(\alpha) \big)\big\}=\mathbf{0},
\end{align}
where $\bX_\alpha$ is the design matrix corresponding to the model $\alpha$. It has been discussed in \cite{LL2010} that a sufficient condition for the uniqueness of the solution to (\ref{e006}) is the combination of Condition \ref{assp2} below and the assumption that $\bX_{\alpha}$ has full rank. In practice, requiring the design matrix $\bX_{\alpha}$ to be full rank is not stringent because a violation means that some explanatory variables can be expressed  as linear combinations of other variables, and thus they can always be eliminated to make the design matrix nonsingular.

For theoretical analysis, we %consider the setting where
assume that the true parameter $\bbeta_0$ is in some sufficiently large, convex, compact set $\mathcal{B}$ in $\mathbf{R}^p$, and that $\|\bbeta^*(\alpha)\|_\infty$ is uniformly bounded by some positive constant for all models $\alpha$ with $|\alpha|\leq K$. Denote by $\bW = (W_1,\cdots, W_n)\t$ where $W_i = Y_i-E[Y_i]$ is the model error for the $i$th observation.
The following conditions are imposed in the theoretical developments of results in  this paper.  %to conduct the asymptotic analysis.

\begin{assumption}\label{assp2} The function
$b(\theta)$ is three times differentiable with $c_0\leq b^{''}(\theta)\leq c_0^{-1}$ and $|b'''(\theta)|\leq c_0^{-1}$ in its domain for some constant $c_0>0$.
\end{assumption}

%\begin{assumption}\label{assp3}
%For each $1\leq i \leq n$, $h_i(\theta)=E[\log f_i(Y_i; \theta, \phi)]$ is smooth in $\theta$, and the differentiation and expectation are exchangeable so that $h_i'(\theta) = E\big[\partial{\log f_i(Y_i; \theta, \phi)}/\partial\theta\big]$ and $h_i''(\theta) = E\big[\partial^2\log f_i(Y_i; \theta, \phi)/\partial\theta^2\big]$.
%\end{assumption}

\begin{assumption}\label{assp4}
For any $\alpha \subset \{1,\cdots, p\}$ such that $|\alpha|\leq K$,  $n^{-1}\bX_\alpha^T\bX_\alpha$ has the smallest and largest eigenvalues bounded from below  and above by $c_1$ and $1/c_1$ for some $c_1>0$, where $K$ is some positive integer satisfying $K >s$ and $K=o(n)$.
\end{assumption}

\begin{assumption}\label{assp5}  For unbounded and non-Gaussian distributed $Y_i$ , there exists a diverging sequence $m_n=o(\sqrt{n})$ such that
\begin{equation}\label{e094}
\sup_{\sbbeta \in \mathcal{B}_1}\max_{1\leq i\leq n}\big|b'\big( |\bx_i\t\bbeta| \big) \big|\leq m_n,
\end{equation}
where $\mathcal{B}_1 = \{\bbeta \in \mathcal{B}: |\supp(\bbeta)|\leq K\}$.  Additionally $W_i$'s follow the uniform sub-Gaussian distribution -- i.e., there exist constants $c_2, c_3 >0$ such that uniformly for all $i=1,\cdots, n$,
\begin{equation}\label{e095}
P(|W_i|\geq t)\leq c_2\exp(-c_3t^2) \text{ for any } t>0.
\end{equation}
\end{assumption}

%We also need to make the following assumption on the design matrix.

\begin{assumption} \label{assp8}
There exits a $\lambda_0 \in [\lambda_{\min}, \lambda_{\max}]$ such that $\alpha_{\lambda_0} = \alpha_0$ and $\|\hbbeta^{\lambda_0} - \bbeta_0\|_2 = O_p(n^{-\pi})$ for $0<\pi < 1/2$. Moreover, for each fixed $\lambda$, $p'_{\lambda}(t)$ is non-increasing over $t\in(0,\infty)$. Also,  $n^\pi\min_{j\in \alpha_0}|\beta_{0j}|\rightarrow \infty$ as $n\rightarrow\infty$.
\end{assumption}

Condition \ref{assp2} implies that the generalized linear model (\ref{eq:lhood}) has smooth and bounded variance function.
It ensures the existence of the Fisher information for statistical inference with model (\ref{eq:lhood}). For commonly used generalized linear models,
Condition \ref{assp2} is satisfied. These include the Gaussian linear model, the logistic regression model, and the Poisson regression model with bounded variance function. Thus, all models fitted in Section \ref{sec:numeric} satisfy this condition.
Condition \ref{assp4} on the design matrix  is important for ensuring the uniqueness of the population parameter $\bbeta^*(\alpha)$.
%{\bf Chengyong: I deleted this sentence because Condition 2 is stronger than the full rank assumption made before.\remove{Similar to our earlier discussion on the design matrix, Condition} \ref{assp4} \remove{is not restrictive for practically applying our method.}
%}
 If a random design matrix is considered, \cite{Wangh2009b} shows that Condition \ref{assp4} holds with probability tending to 1 under appropriate assumptions on the distribution of the predictor vector $\bx_i$,  true model size $s$ and the dimensionality $p$, which are satisfied by the settings in our simulation examples.

 %{\bf Chengyong: could you check if the distribution of design matrix in simulation examples satisfy conditions in Wang(2009)? }

Condition \ref{assp5} is a technical condition used to control the tail behavior of unbounded non-Gaussian response $Y_i$'s.  It is imposed to ensure a general and broad applicability of the method. For many practically applied models such as those in Section \ref{sec:numeric}, this condition is not required.
Inequality (\ref{e094}) is on the mean function of the response variable, while (\ref{e095}) is on the tail probability distribution of the model error. The combination of (\ref{e094}) and (\ref{e095}) controls the magnitude of the response variable $Y_i$ in probability uniformly.
If we further have $\|\bbeta\|_{\infty}\leq C$ with some constant $C>0$ for any $\bbeta\in \mathcal{B}_1$,  with $\mathcal{B}_1$ defined in Condition \ref{assp5}, then
$$
\sup_{\scriptsize{\bbeta}\in \mathcal{B}_1}\max_{1\leq i\leq n}|\bx_i\t\bbeta| \leq \sup_{\scriptsize{\bbeta}\in \mathcal{B}_1}\|\bX_{\scriptsize{\supp(\bbeta)}}\|_{\infty}\|\bbeta\|_\infty \leq CK\max_{ij}|x_{ij}|.
$$
Hence, (\ref{e094}) holds if $|b'(t)|$ is bounded by $m_n$ for all $|t|\leq CK\max_{ij}|x_{ij}|$. Analogous conditions to (\ref{e094}) are made in \cite{FS10} and \cite{BV11} for studying high-dimensional penalized likelihood methods.

%Condition \ref{assp8} is on the regularization method and \add{its validity and practical applicability are supported by}   existing results in literature on variable selection via regurgitation methods.
Since our interest is on tuning parameter selection, we impose Condition \ref{assp8} to ensure that the true model can be recovered by regularization method. All requirements in this condition are not restrictive from the practical perspective, and their validity and applicability can be supported by existing results in literature on variable selection via regularization methods.
Specifically, the first part of
  Condition \ref{assp8} is satisfied automatically if the penalized likelihood method maximizing (\ref{e001}) has the oracle property \citep{FanLi_2001_JASA}.
Meanwhile,  the desirable oracle property and selection consistency for various penalized likelihood methods have been extensively studied recently.
For example,  \cite{ZhaoYu_2006_JMLR} proved that in the linear-model setting, the Lasso method with $l_1$ penalty $p_{\lambda}(t) = \lambda t$ has model selection consistency under the strong irrepresentable condition.  \cite{zhang06} studied the sparsity and bias of the Lasso estimator and established the consistency rate, and
 \cite{lv1} established the weak oracle property of the regularized least-squares estimator with general concave penalty functions.
For generalized linear models, \cite{lv2} proved that the penalized likelihood methods with folded-concave penalty functions enjoy the oracle property in the setting of non-polynomial dimensionality.
The second part of Condition \ref{assp8} is a mild assumption on  $p_{\lambda}(t)$
to avoid excessive bias,  which is satisfied by commonly used penalty functions in practice including those in our numerical examples -- i.e., Lasso, SCAD, and MCP. The last part of Condition \ref{assp8}, $n^\pi\min_{j\in \alpha_0}|\beta_{0j}|\rightarrow \infty$, is a general and reasonable specification on the signal strength for ensuring the model selection sign consistency -- i.e., $\sgn(\hbbeta^{\lambda_0}) = \sgn(\bbeta_0)$, of the estimator $\hbbeta^{\lambda_0}$. This, together with the technical condition $p'_{\lambda_0}(\frac{1}{2}\min_{j\in \alpha_0}|\beta_{0j}|) =o( s^{-1/2}n^{-1/2}a_n^{1/2})$  in Proposition \ref{P4} are used to show that $\|\hbbeta^{\lambda_0}-\hbbeta^*(\alpha_0)\|_2 = o_p({(\log p)^{\xi/2}}/\sqrt{n})$ with $\xi$ defined in Proposition \ref{P3}.
For a more specific data model and penalty function, alternative weaker conditions may replace  Condition \ref{assp8} as long as  the same result holds.

We now establish the uniform convergence of the MLE $\hbbeta^*(\alpha)$ to the population parameter $\bbeta^*(\alpha)$ over all models $\alpha$ with $|\alpha|\leq K$. This intermediate result plays a pivotal role in measuring the goodness of fit of underfitted and overfitted models in Section \ref{sec:proxyGIC}.
\begin{proposition}\label{P1}
Under Conditions \ref{assp2} %, \ref{assp3},
and \ref{assp4},   as $n\to\infty$,
\[
\sup_{|\alpha|\leq K \atop \alpha \subset \{1,\cdots, p\}}\frac{1}{\sqrt{|\alpha|}}\|\hbbeta^*(\alpha) - \bbeta^*(\alpha)\|_2 = O_p\Big(L_n\sqrt{(\log p)/n}\Big),  %\mbox{ when  $S_1$ or $S_2$ is valid. }
\]
 when either a) the $Y_i$'s are bounded or Gaussian distributed, $L_n=O(1)$,   and $\log p=o(n)$; or b)  the $Y_i$'s are unbounded non-Gaussian distributed,  additional Condition  \ref{assp5} holds, $L_n = O(\sqrt{\log n} + m_n)$, and $\log p = o(n/L_n^2)$.
\end{proposition}
Proposition \ref{P1} extends the consistency result of $\hbbeta^*(\alpha_0)$ to $\bbeta_0$ to the uniform setting over all candidate models with model size less than $K$, where there are ${p\choose K} \sim p^{K}$ such models in total. The large amount of candidate models causes the extra term $\log p$ in the convergence rate.

Based on Proposition \ref{P1}, we have the following result on the log-likelihood ratio for non-Gaussian GLM response. It parallels the result (\ref{e018}) in the Gaussian response setting.
\begin{proposition}\label{P3}
Suppose that the design matrix satisfies $\max_{ij}|x_{ij}| = O(n^{\frac{1}{2}-\tau})$ with $\tau \in (0,1/2]$. Then,  under Conditions \ref{assp2} %,  \ref{assp3},
and \ref{assp4}, uniformly for all models $\alpha \supseteq \alpha_0$ with $|\alpha|\leq K$, as $n\to\infty$,
\begin{align*}
   &\ell_n(\hbbeta^*(\alpha))-\ell_n(\bbeta^*(\alpha)) = \frac{1}{2}(\bY-\bmu_0)^T\bH_0^{-1/2} \bB_{\alpha}\bH_0^{-1/2}(\bY-\bmu_0) \\
      &+ |\alpha|^{5/2}O_p(L_n^2n^{\frac{1}{2}-2\tau}(\log p)^{1+\frac{\xi}{2}})+|\alpha|^{4}O_p\big(n^{1-4\tau}(\log p)^{2}\big)+|\alpha|^3O_p\big(L_n^3n^{1-3\tau}(\log p)^{\frac{3}{2}}\big)
\end{align*}
when  a)   $Y_i$'s are bounded, $\xi=1/2$ and $L_n=O(1)$; or b)  $Y_i$'s are unbounded non-Gaussian distributed,  additional Condition \ref{assp5}  holds, $\xi=1$, and $L_n = \sqrt{\log n} + m_n$.
\end{proposition}

\section{Appendix}
\setcounter{equation}{0}

\subsection{Lemmas}

We first present a few lemmas whose proofs are given in the Supplementary Material.

%\subsubsection{Lemma \ref{L3}}
\begin{lemma}\label{L3} Assume $W_1, \cdots, W_n$ are independent and have uniform sub-Gaussian distribution (\ref{e095}). Then, with probability at least $1-o(1)$,
$$
\|\bW\|_\infty \leq  C_1\sqrt{\log n}
$$
with some constant $ C_1>0$. Moreover, for any positive sequence $\tilde L_n \rightarrow \infty$, if $n$ is large enough, there exists some constant $C_2>0$ such that
\[
n^{-1}\sum_{i=1}^n\Big(E\big[W_i\big|\Omega_n\big]\Big)^2 \leq C_2\tilde L_n\exp(-C_2\tilde L_n^2).
\]
\end{lemma}

\begin{lemma}\label{L2} If the $Y_i$'s are unbounded non-Gaussian distributed and Conditions \ref{assp2}-- \ref{assp4} hold, then for any diverging sequence $\gamma_n\rightarrow\infty$ satisfying $\gamma_nL_n\sqrt{K(\log p)/n}\rightarrow 0$,
\begin{align}
\sup_{|\alpha|\leq K}\frac{1}{|\alpha|}Z_{\alpha}\Big(\gamma_nL_n\sqrt{|\alpha|(\log p)/n}\Big)=O_p\big(L_n^2n^{-1}(\log p)\big),
\end{align}
where $L_n=2m_n+C_1\sqrt{\log n}$ with $C_1$ defined in Lemma \ref{L3}. If the $Y_i$'s are bounded and Conditions \ref{assp2}
and \ref{assp4} hold, then the same result holds with $L_n$ replaced with $1$.
\end{lemma}

\begin{lemma}\label{P2} Let $\tilde \bY \equiv (\tilde Y_1, \cdots, \tilde Y_n)\t = \bH_0^{-1/2}(\bY - \bmu_0)$. For any $K =o(n)$,
\begin{align*}
\sup_{\alpha\supset\alpha_0,|\alpha|\leq K}\frac{1}{|\alpha|-|\alpha_0|}\tilde\bY\t (\bB_\alpha-\bB_{\alpha_0})\tilde\bY=O_p\big((\log p)^{\xi}\big),
\end{align*}
where %$\gamma_n \rightarrow \infty$ \footnote{Here $\gamma_n$ seems not needed .}, and
a) $\xi = 1/2$ when the $\tilde Y_i$'s are bounded, and b) $\xi = 1$ when the $\tilde Y_i$'s are uniform sub-Gaussian random variables.
\end{lemma}

We use the empirical process  techniques to prove the main results. We first introduce some notations.  For a given  model $\alpha$ with $|\alpha| \leq K$ and a given $N>0$, define the set
\[
\mathcal{B}_{\alpha}(N)=\{\bbeta\in \mathbf{R}^p: \|\bbeta - \bbeta^*(\alpha)\|_2\leq N,  \supp(\bbeta) =\alpha\}\cup\{\bbeta^*(\alpha)\}.
\]
Consider the negative log-likelihood loss function $\rho(s, Y_i) = -Y_is + b(s)-c(Y_i,\phi)$ for $s\in \mathbf{R}$. Then  $\ell_n(\bbeta)=-\sum_{i=1}^n\rho(\bx_i\t\bbeta,Y_i)$.
Further, define
$Z_{\alpha}(N)$ as
\begin{align}\label{e082}
Z_{\alpha}(N) = \sup_{\bbeta\in \mathcal{B}_{\alpha}(N)}n^{-1}\Big|\ell_{n}(\bbeta)- \ell_{n}(\bbeta^*(\alpha))-E\big[\ell_n(\bbeta)-\ell_n(\bbeta^*(\alpha))\big]\Big|.
\end{align}
 It is seen that $Z_{\alpha}(N)$ is the supreme of the absolute value of an empirical process indexed by $\bbeta \in \mathcal{B}_{\alpha}(N)$. Define the event $\Omega_n = \{\|\bW\|_\infty \leq \tilde L_n\}$ with $\bW = \bY-E[\bY]$  being the error vector and $\tilde L_n$ some positive sequence that  may diverge with $n$. Then, for bounded responses, $P(\Omega_n) = 1$ if $\tilde L_n$ is chosen as a large enough constant;  for unbounded and non-Gaussian responses,  by Lemma \ref{L3}, $P(\Omega_n) = 1-o(1)$ if $\tilde L_n=  C_1\sqrt{\log n}$ with $C_1>0$ a large enough constant. On the event $\Omega_n$, $\|\bY\|_{\infty}\leq m_n + C_1\sqrt{\log n}$. Throughtout, we use $C$ to denote a generic positive constant, and we slightly abuse the notation by using $\bbeta(\alpha)$ to denote either the $p$-vector or its subvector on the support $\alpha$ when there is no confusion.

\subsection{Proof of Proposition \ref{P4}}

\begin{proof}
First note that $\hbbeta_0\equiv \hbbeta^*(\alpha_0)$ maximizes the log-likelihood $\ell_n(\bbeta)$ restricted to model $\alpha_0$. Thus, $\frac{\partial}{\partial \bbeta}\ell_n(\hbbeta_0)=0$. Moreover, it follows from Condition \ref{assp2} that $\frac{\partial}{\partial^2{\scriptsize \bbeta}}\ell_n(\bbeta)=\bX^T\bH(\bbeta)\bX$. Thus, by Taylor's expansion and Condition \ref{assp4} we obtain
\begin{align}
\nonumber 0\geq &\GIC_{a_n}^*(\alpha_0) - \GIC_{a_n}(\lambda_0) = \frac{1}{n}\big(\ell(\hbbeta^{\lambda_0}) - \ell(\hbbeta_0) \big)
 %= \frac{1}{n}(\hbbeta^{\lambda_0} - \hbbeta_0)^T\frac{\partial}{\partial^2\bbeta}\ell(\tbbeta) (\hbbeta^{\lambda_0} - \hbbeta_0)\\&
 \\&=-\frac{1}{n}(\hbbeta^{\lambda_0} - \hbbeta_0)^T\bX^T\bH(\tbbeta)\bX(\hbbeta^{\lambda_0} - \hbbeta_0)\geq -C\|\hbbeta^{\lambda_0}-\hbbeta_0\|_2^2,\label{e052}
\end{align}
where $\tbbeta$ lie on the line segment connecting $\hbbeta^{\lambda_0}$ and $\hbbeta_0$, and we have used $\supp(\hbbeta^{\lambda_0})=\supp(\hbbeta_0)=\alpha_0$  for the last inequality.
It remains to prove that $\|\hbbeta^{\lambda_0} - \hbbeta_0\|_2$ is small.

Let $\hbbeta^{\lambda_0}_{\alpha_0}$ and $\hbbeta_{0,\alpha_0}$ be the subvectors of $\hbbeta^{\lambda_0}$ and $\hbbeta_{0}$ on the support $\alpha_0$, correspondingly. Since $\hbbeta^{\lambda_0}$ minimizes  $\ell_n(\bbeta) + n\sum_{j=1}^np_{\lambda_0}(|\beta_j|)$, it follows from the classical optimization theory that $\hbbeta^{\lambda_0}_{\alpha_0}$  is a critical value, and thus
\begin{align*}
\bX_0^T(\bY - \bb'(\bX_0\hbbeta^{\lambda_0}_{\alpha_0})) + n\bar{p}'_{\lambda_n}(\hbbeta^{\lambda_0}_{\alpha_0})=0,
\end{align*}
where $\bX_{0}$ is the design matrix of the true model, and $\bar{p}'_{\lambda_n}(\hbbeta^{\lambda_0}_{\alpha_0})$ is a vector with components $\sgn(\hbeta^{\lambda_0}_j)p'_{\lambda_0}(|\hbeta^{\lambda_0}_j|)$ and $j\in\alpha_0$. Since $\hbbeta_0$ is the MLE when restricted to the support $\alpha_0$, $\bX_0\t(\bY - \bb'(\bX_0\hbbeta_{0,\alpha_0}))=0$. Thus, the above equation can be rewritten as
 \begin{align}\label{e051}
\bX_0^T\big(\bb'(\bX_0\hbbeta_{0,\alpha_0}) - \bb'(\bX_0\hbbeta^{\lambda_0}_{\alpha_0})\big) + n\bar{p}'_{\lambda_n}(\hbbeta^{\lambda_0}_{\alpha_0})=0,
\end{align}
Now, applying the Taylor's expansion to (\ref{e051}) we obtain that
$
\bX_0^T\bH(\bX_0\bar\bbeta)\bX_0(\hbbeta^{\lambda_0}_{\alpha_0} - \hbbeta_{0,\alpha_0})= n \bar{p}_{\lambda_0}(|\hbbeta^{\lambda_0}_{\alpha_0}|),
$
where $\bar\bbeta$ lies between the line segment connecting $\hbbeta^{\lambda_0}_{\alpha_0}$ and $\hbbeta_{0,\alpha_0}$. Therefore,
\[
\hbbeta^{\lambda_0}_{\alpha_0}-\hbbeta_{0,\alpha_0} = n(\bX_0^T\bH(\bX_0\bar{\bbeta}) \bX_0)^{-1}\bar{p}_{\lambda_0}(\hbbeta^{\lambda_0}_{\alpha_0}).
\]
This together with Conditions \ref{assp2} and \ref{assp4} ensures that
\begin{align}\label{eq001}
\|\hbbeta^{\lambda_0}_{\alpha_0}-\hbbeta_{0,\alpha_0}\|_2 \leq C\|\bar{p}_{\lambda_0}(\hbbeta^{\lambda_0}_{\alpha_0})\|_2.
\end{align}
 Since we have assumed that $\|\hbbeta^{\lambda_0} - \bbeta_0\|_2 = O_p(n^{-\pi})$, it follows that for large enough $n$,
$
\min_{j\in\alpha_0}|\hbeta^{\lambda_0}_j| \geq \min_{j\in\alpha_0}|\beta_{0j}| - n^{-\pi} \geq 2^{-1}\min_{j\in\alpha_0}|\beta_{0j}|
$.
Thus, by theorem assumptions,
$$
\|\bar{p}'_{\lambda_0}(\hbbeta^{\lambda_0}_{\alpha_0})\|_2 \leq \sqrt{s}p'_{\lambda_0}(\frac{1}{2}\min_{ j\in\alpha_0}|\beta_{0j}|) = o(\sqrt{n^{-1}a_n}).
$$
Combing the above inequality with (\ref{eq001}) yields
\begin{align*}
\|\hbbeta^{\lambda_0}-\hbbeta_0\|_2=\|\hbbeta^{\lambda_0}_{\alpha_0}-\hbbeta_{0,\alpha_0}\|_2 \leq o(n^{-1/2}a_n^{1/2}).
\end{align*}
This, together with (\ref{e052}), completes the proof of (\ref{e036}).
%And Theorem \ref{T4} follows correspondingly.

\end{proof}

\subsection{Proof of Proposition \ref{P1}}
\begin{proof}
We first consider the non-Gaussian responses. Using the similar idea in \cite{vandeGeer_2002_Planning}, for a given $N>0$, define a convex combination $\hbbeta_u(\alpha) = u\hbbeta^*(\alpha) + (1-u)\bbeta^*(\alpha)$ with
$
u = (1+\|\hbbeta^*(\alpha) - \bbeta^*(\alpha)\|_2/N)^{-1}.
$
Then, by definition, $\|\hbbeta_u(\alpha) - \bbeta^*(\alpha)\|_2 = u\|\hbbeta^*(\alpha) - \bbeta^*(\alpha)\|_2 \leq N$. If $\supp(\hbbeta_u) \neq\alpha$, then modify the definition of $u$ a little bit by slightly increasing $N$ to make $\supp(\hbbeta_u) = \alpha$. So we assume implicitly that $\supp(\hbbeta_u) = \alpha$ and thus  that $\hbbeta_u\in\mathcal{B}_{\alpha}(N)$. The key is to prove
\begin{align}\label{e034}
\sup_{|\alpha|\leq K}\frac{1}{\sqrt{|\alpha|}}\|\hbbeta_u(\alpha) - \bbeta^*(\alpha)\|_2 = O_p(L_n\sqrt{(\log p)/n}).
\end{align}
Then, by noting that $\|\hbbeta_u(\alpha)-\bbeta^*(\alpha)\|_2 \leq N/2$ implies $\|\hbbeta^*(\alpha) -\bbeta^*(\alpha)\|_2\leq N$, the result in Proposition \ref{P1} is proved.

Now, we proceed to prove (\ref{e034}). By the concavity of the log-likelihood function,
\begin{align*}
\ell_n(\hbbeta_u(\alpha))\geq  u\ell_n(\hbbeta^*(\alpha)) +(1-u)\ell_n(\bbeta^*(\alpha)).
\end{align*}
Since $\hbbeta^*(\alpha)$ maximizes $\ell_n(\bbeta)$ over all models with support  $\alpha$, the above inequality can further be written as $\ell_n(\bbeta^*(\alpha))\leq\ell_n(\hbbeta_u(\alpha)) $.  On the other hand, since $\bbeta^*(\alpha)$ minimizes the KL divergence $I(\bbeta(\alpha))$ in (\ref{e087}), we obtain
\[
E[\ell_n(\bbeta^*(\alpha))-\ell_n(\hbbeta_u(\alpha))]  = I(\hbbeta_u(\alpha)) - I(\bbeta^*(\alpha)) \geq 0,
\]
where $E[\ell_n(\hbbeta_u(\alpha))]=-\sum_{i=1}^nE[\rho(\bx_i\t\hbbeta_u,Y_i)]$ should be understood as $E[\rho(\bx_i\t\hbbeta_u, Y_i)] = \int\rho(\bx_i\t\hbbeta_u,y)dF_i(y)$ with $F_i(\cdot)$ being the distribution function of $Y_i$. Combining these two results yields
\begin{align}\label{e088}
\nonumber0&\leq E[\ell_n(\bbeta^*(\alpha))-\ell_n(\hbbeta_u(\alpha))] \\
&\leq \big(\ell_n(\hbbeta_u(\alpha)) -E[\ell_n(\hbbeta_u(\alpha))] \big)-\big(\ell_n(\bbeta^*(\alpha))-E[\ell_n(\bbeta^*(\alpha))] \big) \leq nZ_{\alpha}(N),
\end{align}
where $Z_{\alpha}(N)$ is defined in (\ref{e082}). On the other hand, by (\ref{e006}), for any $\bbeta(\alpha) \in \mathcal{B}_{\alpha}(N)$,
\begin{align*}
&E[\ell_n(\bbeta(\alpha)) - \ell_n(\bbeta^*(\alpha))] = \bb'(\bX\bbeta_0)\t\bX[\bbeta(\alpha)-\bbeta^*(\alpha)]-\bone\t[\bb(\bX\bbeta(\alpha))-\bb(\bX\bbeta^*(\alpha))]\\
& = \bb'(\bX\bbeta^*(\alpha))\t\bX[\bbeta(\alpha)-\bbeta^*(\alpha)]-\bone\t[\bb(\bX\bbeta(\alpha))-\bb(\bX\bbeta^*(\alpha))]\\
& =-\frac{1}{2}(\bbeta(\alpha)-\bbeta^*(\alpha))\t\bX_{\alpha}\t\tilde\bH\bX_{\alpha}(\bbeta(\alpha)-\bbeta^*(\alpha)),
\end{align*}
where $\tilde\bH  = \text{diag}\big\{\bb''\big(\bX\overline\bbeta(\alpha)\big)\big\}$ and $\overline\bbeta(\alpha)$ lies on the segment connecting $\bbeta^*(\alpha)$ and $\bbeta(\alpha)$. Thus, it follows from Conditions \ref{assp2} and \ref{assp5} that for any $\bbeta(\alpha) \in \mathcal{B}_{\alpha}(N)$,
\[
E\big[\ell_n(\bbeta(\alpha)) - \ell_n(\bbeta^*(\alpha))\big] \leq -\frac{1}{2}c_0c_1n\|\bbeta(\alpha)-\bbeta^*(\alpha)\|_2^2.
\]
This, together with (\ref{e088}), entails that for any $\bbeta(\alpha) \in \mathcal{B}_{\alpha}(N)$,
\[\|\bbeta(\alpha) - \bbeta^*(\alpha)\|_2^2 \leq 2(c_0c_1)^{-1}Z_{\alpha}(N).\]
Since $\hbbeta_u \in \mathcal{B}_\alpha(N)$, taking $N=N_n\equiv \gamma_nL_n\sqrt{|\alpha|(\log p)/n}$ and by Lemma \ref{L2}, we have
\[
\sup_{|\alpha|\leq K}\frac{1}{\sqrt{|\alpha|}}\|\hbbeta_u(\alpha) - \bbeta^*(\alpha)\|_2 \leq 2(c_0c_1)^{-1}\big\{\sup_{|\alpha|\leq K}\frac{1}{|\alpha|}Z_{\alpha}(N_n) \big\}^{1/2}= O_p(L_n\sqrt{(\log p)/n}),
\]
where $L_n=2m_n + O(\sqrt{\log n})$ when the $Y_i$'s are unbounded non-Gaussian, and $L_n = O(1)$ when the $Y_i$'s are bounded. This completes the proof of (\ref{e034}).

 Now consider the Gaussian response. For a given model $\alpha$, we have the explicit form that $\hbbeta^*(\alpha)= (\bX_{\alpha}\t\bX_{\alpha})^{-1}\bX_{\alpha}\t\bY$. Since $\bX_{\alpha}\t(\bX\bbeta^*(\alpha)-\bX\bbeta_0)=0$, direct calculation yields
\[
\hbbeta^*(\alpha) - \bbeta^*(\alpha)  =(\bX_{\alpha}\t\bX_{\alpha})^{-1}\bX_{\alpha}\t\bW.
\]
Since $\bW\sim N(0, \sigma^2I_n)$, it follows that $\hbbeta^*(\alpha) - \bbeta^*(\alpha) \sim N(0,\sigma^2I_{|\alpha|})$. So $\sigma^{-2}\|\hbbeta^*(\alpha) - \bbeta^*(\alpha)\|_2^2\sim \chi_{|\alpha|}^2$. Thus, for $t>0$ there exists $C>0$:
\[
P(\|\hbbeta^*(\alpha) - \bbeta^*(\alpha)\|_2^2 \geq |\alpha|t) \leq C\exp(-C|\alpha|t).
\]
Using a similar method as before,  we obtain that
\[
\sup_{|\alpha|\leq K}|\alpha|^{-1/2}\|\hbbeta^*(\alpha) - \bbeta^*(\alpha)\|_2 = O_p(\sqrt{(\log p)/n} ).
\]
This completes the proof.

\end{proof}

\subsection{Proof of Proposition \ref{P3}}

\begin{proof} %By Proposition \ref{P1}, $\sup_{|\alpha|\leq K}|\alpha|^{-1/2}\|\hbbeta^*(\alpha) - \bbeta^*(\alpha)\|_2 = O_p(L_n\sqrt{(\log p)/n})$ with $L_n$ defined over there.
By Taylor expansion, $\ell_n(\hbbeta^*(\alpha))-\ell_n(\bbeta^*(\alpha)) $ can be written as
\begin{align}\label{e056}
\ell_n(\hbbeta^*(\alpha))-\ell_n(\bbeta^*(\alpha))  &= I_1(\alpha) - I_2(\alpha) + I_3(\alpha), %s_n(\bbeta^*(\alpha))^T(\hbbeta^*(\alpha)-\bbeta^*(\alpha))\\
%&- \frac{1}{2}(\hbbeta^*(\alpha)-\bbeta^*(\alpha))^T\bX_\alpha^T \bH_0\bX_\alpha(\hbbeta^*(\alpha)-\bbeta^*(\alpha))+R_{3,\alpha},
\end{align}
where
\begin{align}
&I_{1}(\alpha) = (\hbbeta^*(\alpha)-\bbeta^*(\alpha))\t\bX^T(\bY - \bb'(\bX\bbeta^*(\alpha))), \label{e049} \\
&I_2(\alpha) = \frac{1}{2}(\hbbeta^*(\alpha)-\bbeta^*(\alpha))^T\bX^T \bH_0\bX(\hbbeta^*(\alpha)-\bbeta^*(\alpha)), \label{e050}
\end{align}
and $I_3(\alpha)$ is the remainder term. We will study them one by one.

We first consider $I_1(\alpha)$.   Since $\hbbeta^*(\alpha)$ is the MLE, it satisfies the first-order equation
$\bX_{\alpha}^T[\bY-\bb'(\bX\hbbeta^*(\alpha))]=0$.
Applying the Taylor expansion to $\bb'(\bX\hbbeta^*(\alpha))$ yields
\begin{align*}
\bX_\alpha^T\bY=\bX_\alpha^T[\bb'(\bX\bbeta^*(\alpha))
+ \bH_0\bX(\hbbeta^*(\alpha)-\bbeta^*(\alpha))+ \bv_\alpha],
\end{align*}
where
\begin{equation}\label{e020}
\bv_\alpha = (v_1,\cdots, v_n)^T \text{ with } v_i=\frac{1}{2}b'''(\bx_{i}^T\tbbeta^*(\alpha))\big(\bx_{i}^T(\hbbeta^*(\alpha)-\bbeta^*(\alpha))\big)^2
\end{equation}
and $\tbbeta^*(\alpha)$ lying on the line segment connecting $\bbeta^*(\alpha)$ and $\hbbeta^*(\alpha)$. Since (\ref{e006}) ensures $\bX_{\alpha}^T\bb'(\bX\bbeta_0) = \bX_{\alpha}^T\bb'(\bX\bbeta^*(\alpha))$.  Thus we have
\begin{align}\label{e046}
\nonumber\hbbeta^*(\alpha)-\bbeta^*(\alpha)&=(\bX_\alpha^T\bH_0\bX_{\alpha})^{-1}\bX_\alpha^T(\bY-\bb'(\bX\bbeta^*(\alpha))
- \bv_\alpha)\\
&=(\bX_\alpha^T\bH_0\bX_{\alpha})^{-1}\bX_\alpha^T(\bY-\bmu_0-\bv_\alpha).
\end{align}
Combining (\ref{e049}) and (\ref{e046}), we obtain
\begin{align}
I_1(\alpha)=(\bY-\bmu_0)^T\bH_0^{-1/2} \bB_{\alpha}\bH_0^{-1/2}(\bY-\bmu_0)+R_{1,\alpha}, \label{e025}
\end{align}
where $\bB_\alpha$ is defined in (\ref{e055})  and $R_{1,\alpha} = -\bv_\alpha^T\bH_0^{-1/2}\bB_{\alpha}\bH_0^{-1/2}\bW$. We only need to study $R_{1,\alpha}$. By the Cauchy-Schwartz inequality, we have
\begin{align}\label{e027}
|R_{1,\alpha}|& \leq \|\bB_{\alpha}\bH_0^{-1/2}\bW\|_2\|\bH_0^{-1/2}\bv_\alpha\|_2\leq \Big(\|\bB_{\alpha_0} \bH_0^{-1/2}\bW\|_2+\|\tilde R_{1,\alpha}\|_2\Big)\|\bH_0^{-1/2}\bv_\alpha\|_2,
\end{align}
where $
\tilde{R}_{1,\alpha} = (\bB_\alpha-\bB_{\alpha_0})\bH_0^{-1/2}\bW$. We consider the terms on the very right-hand side of (\ref{e027}) one by one.
By the Markov's inequality, and noting that $\bH_0 = E[\bW\bW\t]$ and $\tr(\bB_{\alpha_0}\bB_{\alpha_0})=|\alpha_0|$,  we can derive that for any $\gamma_n\rightarrow\infty$
\begin{align*}
&P\Big(\|\bB_{\alpha_0} \bH_0^{-1/2}\bW\|_2\geq \sqrt{|\alpha_0|\gamma_n}\Big)\leq \frac{1}{|\alpha_0|\gamma_n}E[\|\bB_{\alpha_0} \bH_0^{-1/2}\bW\|_2^2]\\
& =\frac{1}{|\alpha_0|\gamma_n}\text{tr}\{\bB_{\alpha_0}\bH_0^{-1/2} E[\bW\bW^T]\bH_0^{-1/2}\bB_{\alpha_0}\} = \frac{1}{\gamma_n} \rightarrow 0.
\end{align*}
Therefore,
\begin{align}\label{e045}
\|\bB_{\alpha_0} \bH_0^{-1/2}\bW\|_2 = O_p(\sqrt{|\alpha_0|}).
\end{align}
Next, by Lemma \ref{P2} we obtain that uniformly for all $\alpha$:
\begin{align}\label{e042}
\{|\alpha|-|\alpha_0|\}^{-1/2}\|\tilde{R}_{1,\alpha}\|_2 = O_p((\log p)^{\xi/2}),
\end{align}
where $\xi$ is defined therein. Finally we consider $\|\bH_0^{-1/2}\bv_\alpha\|_2$. Since $b'''(\cdot)$ is bounded, $\max_{ij}|x_{ij}|=O(n^{\frac{1}{2}-\tau})$ and $\supp(\hbbeta^*(\alpha))=\supp(\bbeta^*(\alpha))=\alpha$, by (\ref{e020}) and Condition \ref{assp4},
\begin{align}
\nonumber&\|\bH_0^{-1/2}\bv_\alpha\|_2\leq C\|\bv_\alpha\|_2\leq C\Big(\sum_{i=1}^n|\bx_{i}^T(\hbbeta^*(\alpha)-\bbeta^*(\alpha))|^4\Big)^{1/2} \\
&\leq C|\alpha|n^{\frac{3}{2}-2\tau}\|\hbbeta^*(\alpha)-\bbeta^*(\alpha)\|_2^2= |\alpha|^2O_p(L_n^2n^{\frac{1}{2}-2\tau}(\log p)). \label{e029}
\end{align}
Combining (\ref{e027}) -- (\ref{e029}), and in view of (\ref{e025}),  we obtain that
\begin{align} \label{e032}
 I_1(\alpha) = (\bY-\bmu_0)^T\bH_0^{-1/2} \bB_{\alpha}\bH_0^{-1/2}(\bY-\bmu_0)+R_{1,\alpha},
 \end{align}
where uniformly for all overfitted models $\alpha$,
\begin{align}\label{e030}
R_{1,\alpha} = |\alpha|^{5/2}O_p(L_n^2n^{\frac{1}{2}-2\tau}(\log p)^{1+\frac{\xi}{2}}).
\end{align}

Next,  we consider $I_2(\alpha)$ defined in (\ref{e050}). By (\ref{e046}) we have the decomposition
\begin{align}
\nonumber I_2(\alpha) & = \frac{1}{2}[\hbbeta^*(\alpha)-\bbeta^*(\alpha)]^T\bX_\alpha^T\bH_0\bX_{\alpha}[\hbbeta^*(\alpha)-\bbeta^*(\alpha)]\\
&=\frac{1}{2}(\bY-\bmu_0)^T\bH_0^{-1/2} \bB_{\alpha}\bH_0^{-1/2}(\bY-\bmu_0)+ \frac{1}{2}R_{2,\alpha}-R_{1,\alpha}, \label{e026}
\end{align}
where $R_{2,\alpha} = \bv^T_\alpha\bH_0^{-1/2} \bB_{\alpha}\bH_0^{-1/2}\bv_\alpha$, and $R_{1,\alpha}$ is defined in (\ref{e032}) and (\ref{e030}). We only need to study $R_{2,\alpha}$. Since $b''(\cdot)$ is bounded, $\max_{i,j}|x_{ij}|=O(n^{\frac{1}{2}-\tau})$ and $\bB_{\alpha}$ is a projection matrix, it is easy to derive that
\begin{align}
R_{2,\alpha}& = \bv^T_\alpha\bH_0^{-1/2} \bB_{\alpha}\bH_0^{-1/2}\bv_\alpha
\leq\bv^T_\alpha\bH_0^{-1}\bv_{\alpha}\leq C \|\bv_\alpha\|_2^2=|\alpha|^{4}O_p\big(n^{1-4\tau}(\log p)^{2}\big), \label{e031}
\end{align}
where the last step is because of Theorem \ref{T1} and (\ref{e020}). The above result is uniformly over all $\alpha$ with $|\alpha|\leq K$. This, together with (\ref{e050}), and (\ref{e030})--(\ref{e031}) ensures that uniformly for all overfitted models $\alpha$,
\begin{align}\label{e047}
\nonumber I_2(\alpha)=& \frac{1}{2}(\bY-\bmu_0)^T\bH_0^{-1/2} \bB_{\alpha}\bH_0^{-1/2}(\bY-\bmu_0)\\
 &+ |\alpha|^{5/2}O_p(L_n^2n^{\frac{1}{2}-2\tau}(\log p)^{1+\frac{\xi}{2}})+|\alpha|^{4}O_p\big(n^{1-4\tau}(\log p)^{2}\big).
\end{align}

Finally, we consider $I_{3}(\alpha)$ in (\ref{e056}).  Since $b'''(\cdot)$ is bounded, by Theorem \ref{T1} we have
\begin{align*}
|I_{3}(\alpha)|\leq Cn^{\frac{5}{2}-3\tau}|\alpha|^{3/2}\|\hbbeta^*(\alpha)-\bbeta^*(\alpha)\|_2^3 = |\alpha|^3O_p\big(n^{1-3\tau}L_n^3(\log p)^{\frac{3}{2}}\big),
\end{align*}
where this result is uniformly over all $\alpha$ with $|\alpha|\leq K$. The above result, together with (\ref{e032}), (\ref{e047}) and (\ref{e056}), completes the proof of Proposition \ref{P3}.

\end{proof}

%\subsection{Goodness of Fit for All Models}

\subsection{Proof of Theorem \ref{C1}}

%Since the scaled deviance measure is the difference of log-likelihood functions, we first study the asymptotic properties of the log-likelihood in the following two theorems.
\begin{proof}
We note that Theorem \ref{C1} is a direct consequence of the following two propositions, the proofs of which are  given in the Supplementary Material.

\begin{proposition}\label{T1}
 In either situation a) or b) in Proposition \ref{P1}, and under the same conditions,   as $n\rightarrow \infty$,
 \begin{align*}
\sup_{|\alpha|\leq K  \atop \alpha \subset \{1,\cdots, p\}}\frac{1}{n|\alpha|}\Big(\ell_n( \hbmu_\alpha^*; \bY)-\ell_n(\bmu_{\alpha}^*; \bY)\Big) =O_p(n^{-1}L_n^2\log p).
\end{align*}
\end{proposition}

\begin{proposition}\label{T5}
Under Conditions \ref{assp2}%, \ref{assp3},
and \ref{assp4}, as $n\rightarrow \infty$,
\[
\sup_{|\alpha|\leq K  \atop \alpha \subset \{1,\cdots, p\}}\frac{1}{n|\alpha|}\big|\ell_n(\bmu_{\alpha}^*;\bY) - E[\ell_n(\bmu_{\alpha}^*;\bY)]\big| = O_p(\sqrt{(\log p)/n})
\]
when either a) the $Y_i$'s are bounded or Gaussian distributed and $\log p = o(n)$; or b) the $Y_i$'s are unbounded and non-Gaussian distributed, additional Condition \ref{assp5} holds, $\max_{ij}|x_{ij}| = O(n^{\frac{1}{2}-\tau})$ with $\tau \in (0,1/2]$, and $K^2\log p = o(n^{2\tau})$.
\end{proposition}
\end{proof}
%The first proposition is a consequence of Theorem  \ref{P1}, and the proof of Proposition \ref{T5} is given in the Supplementary Material.  %It shows that  the log-likelihood $\ell_n(\bmu_{\alpha}^*; \bY)$ can be uniformly well approximated by its estimate $\ell_n(\hbmu_{\alpha}^*; \bY)$  with $ \hbmu_{\alpha}^* = \bb'(\bX\hbbeta^*(\alpha))$.

%Additionally, we show in the next theorem that the population quantity  $E[\ell_n(\bmu_{\alpha}^*;\bY)]$ can be approximated uniformly well  by its empirical quantity $\ell_n(\bmu_{\alpha}^*;\bY)$.

%Now we are ready to characterize the scaled deviance measure.

\subsection{Proofs of Theorem \ref{T2}}

\begin{proof}
 Theorem \ref{T2} follows directly from Lemma \ref{P2} and Proposition \ref{P3}.
\end{proof}

\subsection{Proofs of Theorem \ref{T4}}

\begin{proof}
Combining (\ref{e013}) with (\ref{e021}), and in view of Theorems \ref{T1} and \ref{T2}, we obtain that if $\delta_nK^{-1} R_n^{-1} \rightarrow \infty$,  $a_n$ satisfies
$n \delta_n s^{-1} a_n^{-1}\to \infty$,  and $a_n \psi_n^{-1}\to \infty$, then
\begin{align}\label{e022}
P\left(\inf_{\alpha\supsetneq\alpha_0}\GIC^*_{a_n}(\alpha)-\GIC^*_{a_n}(\alpha_0) >\frac{\delta_n}{2} \text{ and } \inf_{\alpha\not\supset\alpha_0}\GIC^*_{a_n}(\alpha)-\GIC^*_{a_n}(\alpha_0)>\frac{a_n}{2n} \right) \longrightarrow 1,
\end{align}
where $R_n$ and $\psi_n$ are specified in Theorems \ref{C1} and \ref{T2}. This, together with Proposition \ref{P4} and (\ref{e053}), completes the proof of the theorem.
\end{proof}

%\bibliography{f:/text/reference/au}

\renewcommand{\baselinestretch}{1.}

\newpage

%\end{document}
\setcounter{page}{1}
\setcounter{section}{0}
\setcounter{equation}{0}

\renewcommand{\theequation}{A.\arabic{equation}}
\setcounter{equation}{0}

\begin{center}{\bf \large Supplementary Material to ``Tuning Parameter Selection in High-Dimensional Penalized Likelihood''}

\bigskip

{\bf Yingying Fan and Cheng Yong Tang}
\end{center}

This Supplementary Material contains proofs of Lemmas 1-3, and Propositions 4 and 5.

\section{Proofs of Lemmas}

\subsection{Proof of Lemma \ref{L3}}

%\subsubsection{Lemma \ref{L3}}
%\begin{lemma}\label{L3} Assume $W_1, \cdots, W_n$ are independent and have uniform sub-Gaussian distribution (\ref{e095}). Then with probability at least $1-o(1)$,
%$$
%\|\bW\|_\infty \leq  C_1\sqrt{\log n}
%$$
%with some constant $ C_1>0$. Moreover, for any positive sequence $\tilde L_n \rightarrow \infty$, if $n$ is large enough, there exists come constant $C_2>0$ such that
%\[
%n^{-1}\sum_{i=1}^n\Big(E\big[W_i\big|\Omega_n\big]\Big)^2 \leq C_2\tilde L_n\exp(-C_2\tilde L_n^2).
%\]
%\end{lemma}

\begin{proof}
The first result follows trivially from the definition of the uniform sub-Gaussian distribution. So, we only prove the second result.

Since $E[W_i] =0$, by the definition of condition expectation,
\begin{equation}\label{e004}
E\big[W_i\big| |W_i|\leq  \tilde L_n\big] = \frac{E[W_i1\{|W_i|\leq \tilde L_n\}]}{P(|W_i|\leq \tilde L_n)}=-\frac{1}{P(|W_i|\leq \tilde L_n)}E[W_i1\{|W_i|> \tilde L_n\}].
\end{equation}

Next, note that $|E[W_i1\{|W_i|> \tilde L_n\}]|\leq E[|W_i|1\{|W_i| > \tilde L_n\}]$. By the definition of expectation, the right-hand side can be further bounded as
\begin{align*}
E\big[|W_i|1\{|W_i|> \tilde L_n\} \big]& = \int_{0}^{\infty}P\big(|W_i|1\{|W_i|> \tilde L_n\}\geq t\big)dt\\&= \tilde L_nP(|W_i|>\tilde L_n) + \int_{\tilde L_n}^{\infty}P( |W_i|> t)dt.
\end{align*}
Further, by (\ref{e095}) in Condition \ref{assp5} and the tail inequality for Gaussian density $\int_{\tilde L_n}^{\infty}\exp(-c_3t^2)dt \leq C\tilde L_n^{-1}\exp(-C\tilde L_n^2)$, it follows that
\begin{align*}
&\big|E[W_i1\{|W_i|> \tilde L_n\}]\big| %\leq \tilde L_nP(|W_n|> \tilde L_n) + \int_{\tilde L_n}^{\infty}P(|W_i|>t)dt\\
\leq c_2\tilde L_n\exp(-c_3\tilde L_n^2)+c_2\int_{\tilde L_n}^{\infty}\exp(-c_3t^2)dt\leq C\tilde L_n\exp(-C\tilde L_n^2).
\end{align*}
This, together with  (\ref{e004}) and (\ref{e095}), yields
\begin{align*}
E\big[W_i\big| |W_i|\leq  \tilde L_n\big]\leq C\tilde L_n\exp(-C\tilde L_n^2).
\end{align*}
This and the independence of $W_i$ ensure the second result in the lemma.
\end{proof}

%\subsubsection{Lemma \ref{L2}}

%begin{lemma}\label{L2} If $Y_i$'s are unbounded non-Gaussian distributed and Conditions \ref{assp2}-- \ref{assp4} hold, then for any diverging sequence $\gamma_n\rightarrow\infty$ satisfying $\gamma_nL_n\sqrt{K(\log p)/n}\rightarrow 0$,
%\begin{align}
%\sup_{|\alpha|\leq K}\frac{1}{|\alpha|}Z_{\alpha}\Big(\gamma_nL_n\sqrt{|\alpha|(\log %p)/n}\Big)=O_p\big(L_n^2n^{-1}(\log p)\big),
%\end{align}
%where $L_n=2m_n+C_1\sqrt{\log n}$ with $C_1$ defined in Lemma \ref{L3}. If $Y_i$'s are bounded and %Conditions \ref{assp2}, \ref{assp3}, and \ref{assp4} hold, then the same result holds with $L_n$ replaced %with $1$.
%\end{lemma}

\subsection{Proof of Lemma \ref{L2}}

\begin{proof}
Define $\tilde Z_{\alpha}(N)$ as
\begin{align}\label{e083}
\tilde Z_{\alpha}(N) = \sup_{\bbeta\in \mathcal{B}_{\alpha}(N)}n^{-1}\Big|\ell_{n}(\bbeta)- \ell_{n}(\bbeta^*(\alpha))-E\big[\ell_n(\bbeta)-\ell_n(\bbeta^*(\alpha))|\Omega_n\big]\Big|.
\end{align}
%where $E[\ell_n(\bbeta)|\Omega_n] = -\sum_{i=1}^nE[\rho(\bx_i\t\bbeta,Y_i|\Omega_n)]$ with $E[\rho(\bx_i\t\bbeta, Y_i|\Omega_n)] = \int\rho(\bx_i\t\bbeta,y)dF_i(y\big| |Y_i|\leq \tilde L_n)$ and $F_i(\cdot \big| |Y_i|\leq \tilde L_n)$ the conditional distribution function of $Y_i$.
By the definition of $Z_{\alpha}(N)$ and $\tilde Z_{\alpha}(N)$, we have the following triangular inequality:
\begin{align}\label{e005}
\sup_{|\alpha|\leq K}\frac{1}{|\alpha|}Z_{\alpha}(N)\leq \sup_{|\alpha|\leq K}\frac{1}{|\alpha|}\tilde Z_{\alpha}(N) + \sup_{|\alpha|\leq K, \bbeta\in \mathcal{B}_{\alpha}(N)}\frac{1}{|\alpha|}R_{\alpha}(\bbeta) \equiv I_1(N)+I_2(N),
\end{align}
where $R_{\alpha}(\bbeta)=\frac{1}{n}\big|(E[\bY] -E[\bY|\Omega_n])\t\bX(\bbeta-\bbeta^*(\alpha))\big|$. We will prove that with $N=\gamma_nL_n\sqrt{|\alpha|(\log p)/n}$,
\begin{align}
&I_1(N) = o((\log p)/n), \label{e007}\\
&
I_2(N)= O_p\big(L_n^2n^{-1}(\log p)\big).\label{e009}
\end{align}
Then combining (\ref{e005})-(\ref{e009}) completes the proof.

We now proceed to prove (\ref{e007}). Note that $n^{-1}\bX_{\alpha}\t\bX_{\alpha}$ has bounded eigenvalues as assumed in Condition \ref{assp4}, by Cauchy-Schwarz inequality and Lemma \ref{L3} we have
\begin{align*}
&R_{\alpha}(\bbeta)\leq n^{-1}\|E[\bW|\Omega_n]\|_2\|\bX_{\alpha}(\bbeta-\bbeta^*(\alpha)\|_2 \\
&\leq  c_1^{-1/2}\|n^{-1/2}E[\bW|\Omega_n]\|_2\|\bbeta-\bbeta^*(\alpha)\|_2 \leq C\exp(-C\tilde L_n^2)N.
\end{align*}
Taking $\tilde L_n = C_1\sqrt{\log n}$ with $C_1$ being some large positive constant completes the proof of (\ref{e007}).

Now, we prove (\ref{e009}). The key is to use concentration inequalities in the empirical process literature to prove the uniform convergence result for $\tilde Z_{\alpha}(N)$. For $\bbeta_1, \bbeta_2\in \mathcal{B}_{\alpha}(N)$, by the middle-value theorem $b(\bx_i\t\bbeta_1) - b(\bx_i\t\bbeta_2) = b'(\bx_i\t\tbbeta)(\bx_i\t\bbeta_1 - \bx_i\t\bbeta_2)$ with $\tbbeta$ lying on the line segment connecting $\bbeta_1$ and $\bbeta_2$. Moreover, it follows from (\ref{e094}) in Condition \ref{assp5} that  $|b'(\bx_i\t\tbbeta)| \leq m_n$.  Thus, conditioning on $\Omega_n$, $\rho(\cdot,Y_i)$ satisfies the Lipschitz inequality
\begin{align}\label{e085}
\nonumber |\rho(\bx_i\t\bbeta_1,Y_i)-\rho(\bx_i\t\bbeta_2, Y_i)| &= |-Y_i(\bx_i\t\bbeta_1 - \bx_i\t\bbeta_2)+b(\bx_i\t\bbeta_1)-b(\bx_i\t\bbeta_2)| \\
&\leq L_n|\bx_i\t(\bbeta_1 - \bbeta_2)|
\end{align}
with $L_n =\tilde L_n + 2m_n$. %For bounded responses, both $\tilde L_n$ and $m_n$ can be replaced with the upper bound of $Y_i$'s, and thus $L_n$ can be replaced with some large enough positive constant independent of $n$.

Let $\veps_1, \cdots, \veps_n$ be a Rademacher sequence, independent of $W_1, \cdots, W_n$. By the symmetrization theorem combined with the Lipschitz condition (\ref{e085}) and the concentration inequality (see, for example, Theorems 14.3 and 14.4 in \cite{BV11}), we obtain that
\begin{align}\label{e081}
\nonumber E[\tilde Z_{\alpha}(N)| \Omega_n] &\leq 2 E\Big[\sup_{\bbeta\in \mathcal{B}_{\alpha}(N)}n^{-1}\big|\sum_{i=1}^n\veps_i\big(\rho(\bx_i\t\bbeta, Y_i) - \rho(\bx_i\t\bbeta^*(\alpha), Y_i) \big)\big|\Omega_n\Big] \\
&\leq 4L_nE\Big[\sup_{\bbeta\in \mathcal{B}_{\alpha}(N)}n^{-1}|\sum_{i=1}^n\veps_i\big(\bx_i\t\bbeta - \bx_i\t\bbeta^*(\alpha)\big)\Big].
\end{align}
Now, by the Cauchy-Schwartz inequality,
\begin{equation}\label{e090}
\sup_{\bbeta\in \mathcal{B}_{\alpha}(N)}\big|\frac{1}{n}\sum_{i=1}^n\veps_i\big(\bx_i\t\bbeta - \bx_i\t\bbeta^*(\alpha)\big)\big|\leq  \Big(\sup_{\bbeta\in \mathcal{B}_{\alpha}(N)}\|\bbeta - \bbeta^*(\alpha)\|_2 \Big)\Big(\sum_{j\in \alpha}\big|\frac{1}{n}\sum_{i=1}^n(\veps_ix_{ij})^2\big|\Big)^{1/2}.
\end{equation}
Since $\sum_{i=1}^nx_{ij}^2 = n$ for  any $j \in \{1,\cdots, p\}$, it follows from the definition of $\veps_i$'s that
 \begin{equation}\label{e089}
 E\Big(\sum_{j\in \alpha}\big|n^{-2}\sum_{i=1}^n(\veps_ix_{ij})^2\big|\Big)^{1/2}\leq \Big(n^{-2}\sum_{j\in \alpha}\sum_{i=1}^n E[(\veps_ix_{ij})^2] \Big)^{1/2} = \sqrt{|\alpha|/n}.
 \end{equation}
 Combing (\ref{e081})--(\ref{e089}) ensures that
\begin{equation}\label{e058}
E[\tilde Z_{\alpha}(N)| \Omega_n]  \leq 4L_nN\sqrt{|\alpha|/n}.
\end{equation}

For all $|\alpha|\leq K$, by Condition \ref{assp4},
\[n^{-1}\sum_{i=1}^n\big(L_n\bx_i\t(\bbeta(\alpha) - \bbeta_0)\big)^2 = n^{-1} L_n^2(\bbeta(\alpha)-\bbeta_0)\t\bX_{\alpha}\t\bX_{\alpha}(\bbeta(\alpha)-\bbeta_0)\leq c_1^{-1}L_n^2N^2.\]
Combining this with the Lipschitz inequality (\ref{e085}), and applying the  Massart's concentration theory (see Theorem 14.2 in \cite{BV11}) yields that for any $t>0$,
\[
P\Big(\tilde Z_{\alpha}(N) \geq E[\tilde Z_{\alpha}(N)|\Omega_n] + t \Big| \Omega_n\Big)\leq \exp(-nc_1t^2/(2L_n^2N^2)).
\]
Taking $t = 4NL_nu\sqrt{|\alpha|/n}$ with $u>0$, and in view of the bound (\ref{e081}), we obtain
\begin{equation*}
P\Big(\tilde Z_{\alpha}(N) \geq 4L_nN\sqrt{|\alpha|/n}(1+u) \big| \Omega_n\Big)\leq \exp(-8c_1|\alpha|u^2).
\end{equation*}
Further note that ${p\choose k}\leq (pe/k)^k$ for any $0\leq k\leq p$. So taking $N = N_n \equiv L_n\sqrt{|\alpha|/n}(1+u)$, we have
\begin{align}\label{e015}
\nonumber &P\Big(\sup_{|\alpha|\leq K}\frac{1}{|\alpha|}\tilde Z_{\alpha}(N_n) \geq 4L_n^2n^{-1}(1+ u)^2 \big| \Omega_n\Big)\\
&\leq \sum_{|\alpha|\leq K} P\Big(\tilde Z_{\alpha}(N_n) \geq 4|\alpha|L_n^2n^{-1}(1+ u)^2 \big| \Omega_n\Big)
\leq \sum_{k\leq K}(pe/k)^k\exp(-8c_1ku^2).
\end{align}
Now, taking $u = \gamma_n\sqrt{\log p}$ with $\gamma_n$ some slowly diverging sequence in (\ref{e015}), we have
\begin{equation*}
P\Big(\sup_{|\alpha|\leq K}\frac{1}{|\alpha|}\tilde Z_{\alpha}(N_n) \geq 4\gamma_n^2L_n^2n^{-1}\log p \big| \Omega_n\Big) \rightarrow 0.
\end{equation*}
Finally, since for any event $A$ we have $P(A) \leq P(A|\Omega_n) + P(\Omega_n^c)$, it follows from the result above that
\begin{align*}
&P\Big(\sup_{|\alpha|\leq K}\frac{1}{|\alpha|}\tilde Z_{\alpha}(N_n)\geq 4\gamma_n^2 L_n^2n^{-1}\log p\Big) \leq o(1)+P(\Omega_n^c).
\end{align*}
By Lemma \ref{L3}, $P(\Omega_n^c)\rightarrow 0$ if $\tilde L_n = C_1\sqrt{\log n}$ with $C_1$ being a large enough constant. Thus, (\ref{e009}) has been proved and the result in the lemma follows.

If in addition the $Y_i$'s are bounded, then $\tilde L_n$ and $m_n$ can both be taken as the upper bound of the $Y_i$'s, and $\Omega_n$ becomes the whole probability space. Thus, the second result in Lemma \ref{L3} follows easily by similar arguments.
\end{proof}

%\subsubsection{Lemma \ref{L4}}

%\subsubsection{Lemma \ref{P2}}

%\begin{lemma}\label{P2} Let $\tilde \bY \equiv (\tilde Y_1, \cdots, \tilde Y_n)\t = \bH_0^{-1/2}(\bY - %\bmu_0)$. For any $K =o(n)$,
%\begin{align*}
%\sup_{\alpha\supset\alpha_0,|\alpha|\leq K}\frac{1}{|\alpha|-|\alpha_0|}\tilde\bY\t (\bB_\alpha-\bB_{\alpha_0})\tilde\bY=O_p\big((\log p)^{\xi}\big),
%\end{align*}
%where %$\gamma_n \rightarrow \infty$ \footnote{Here $\gamma_n$ seems not needed .}, and
%a) $\xi = 1/2$ when $\tilde Y_i$'s are bounded, b) $\xi = 1$ when $\tilde Y_i$'s are uniform sub-Gaussian random variables.
%\end{lemma}

\subsection{Proof of Lemma \ref{P2}}
\begin{proof} To ease the presentation, denote by $k=|\alpha|-|\alpha_0|$, and $\bP_\alpha=(P_{ij})=\bB_\alpha - \bB_{\alpha_0}$. Then, $\bP_{\alpha}$ is a projection matrix. Since $\text{tr}(\bP_\alpha-\bP_{\alpha_0})=k$ and $\text{tr}\big((\bP_\alpha)^2\big)=\text{tr}(\bP_\alpha)=k$, it is easy to obtain that
$
\sum_{i=1}^n P_{ii}=k \text{ and } \sum_{i, j} P_{ij}^2=k.
$
Moreover, since $\bP_\alpha$ is the projection matrix, it follows that $0\leq P_{ii}\leq 1$. The key is the following decomposition:
\begin{align}\label{e033}
\frac{1}{k}\tilde\bY^T\bP_\alpha\tilde\bY = \frac{1}{k}\sum_{i=1}^n P_{ii}\tilde Y_i^2 + \frac{1}{k}\sum_{i\neq j}P_{ij}\tilde Y_i\tilde Y_j \equiv I_1(\alpha) + I_2(\alpha).
\end{align}
The first term $I_1(\alpha)$ is a summation of independent random variables, and its tail probability has been thoroughly studied in the literature. The difficulty comes from the second term $I_2(\alpha)$, whose summands are not independent. To overcome this difficulty, we use the decoupling inequality. According to \cite{Pena1994}, the tail probability of $I_2(\alpha)$ can be obtained by comparing with the random variable $\tilde I_2(\alpha) = k^{-1}\sum_{i\neq j}P_{ij}\tilde Y_i \tilde Y_j^*$, where $(\tilde Y_1^*,\cdots, \tilde Y_n^*)$ is an independent copy of $(\tilde Y_1,\cdots, \tilde Y_n)$. Specifically, there exists a constant $C>0$ independent of $n$ and $\alpha$, such that
\begin{align}\label{e012}
P\Big(k^{-1}|\sum_{i\neq j}P_{ij}\tilde Y_i\tilde Y_j| \geq t\Big) \leq CP\Big(k^{-1}|\sum_{i\neq j}P_{ij}\tilde Y_i\tilde Y_j^*| \geq C^{-1}t\Big),
\end{align}
where the right-hand side is the tail probability of the sum of independent random variables.

We separate the cases when $Y_i$'s are bounded or sub-Gaussian.

\smallskip

{\noindent\bf When the $Y_i$'s are bounded}:
We first consider $I_1(\alpha)$. Note that $E[I_1(\alpha)] = 1$ and $\sum_{i=1}^n k^{-2}P_{ii}^2\leq k^{-2}\sum_{i, j}^n P_{ij}^2 = k^{-1}$. Since the $\tilde Y_i$'s are independent, by Hoeffding's inequality (see \citeauthor{BV11}, \citeyear{BV11}),   we obtain that for any $x>0$,
\begin{align*}
P\left(I_1(\alpha)\geq 1+x \right) \leq 2\exp\big(-\frac{Cx^2}{\sum_{i=1}^n k^{-2}P_{ii}^2}\big)\leq 2\exp\Big(-Ckx^2\Big).
\end{align*}
Thus, taking $x=\sqrt{\gamma_n\log p}$ with $\gamma_n$ any diverging sequence and noting that ${p \choose k} \leq (pe/k)^k$ for any  positive integers $p,k$, we have
\begin{align}
\nonumber&P\Big(\sup_{|\alpha|\leq K}I_1(\alpha)\geq 1+\sqrt{\gamma_n\log p} \Big)\leq \sum_{|\alpha|\leq K}P\Big(I_1(\alpha)\geq 1 + \sqrt{\gamma_n\log p}\Big) \\
 \leq &2\sum_{k=1}^{K-s}{p-s \choose k}\exp(-Ck\gamma_n\log p)
\leq 2 C\sum_{k=1}^{ K}((p-s)e/k)^k\exp(-Ck\gamma_n\log p)\rightarrow 0. \label{e014}
\end{align}
This ensures that
\begin{align}\label{e008}
\sup_{|\alpha|\leq K}I_1(\alpha) = O_p(\sqrt{\log p}).
\end{align}

Next, we consider $I_2(\alpha)$. Since $\sum_{i\neq j} P_{ij}^2 = \sum_i(P_{ii}-P_{ii}^2) < k$, by (\ref{e012}) and  the Hoeffding's inequality
\begin{align*}
&P(|I_2(\alpha)|\geq t) \leq CP\Big(\frac{1}{k}|\sum_{i\neq j}P_{ij}\tilde Y_i\tilde Y_j^*| \geq C^{-1}t\Big) \leq C\exp\Big(-\frac{C^{-2}k^2t^2}{\sum_{i\neq}P_{ii}^2}\Big) \leq C\exp(-Ckt^2).
\end{align*}
Thus, using a  similar argument as that for $I_1(\alpha)$ we obtain
\begin{align}\label{e017}
\sup_{|\alpha|\leq K}I_2(\alpha) = O_p(\sqrt{\log p}).
\end{align}
Combining (\ref{e008}) and (\ref{e017}) with (\ref{e033}) completes the proof.

\smallskip

{\noindent\bf When $Y_i$'s are sub-Gaussian}: First consider $I_1(\alpha)$. It follows easily from Conditions \ref{assp2} and \ref{assp5} that $\tilde Y_i$'s are also sub-Gaussian. By Condition \ref{assp5} and Stirling's formula $m!\sim (2\pi m)^{1/2}(m/e)^m$, we have for all $m \geq 2$,
\begin{align}\label{e057}
\nonumber E|\tilde Y_i^2|^m &= m\int x^{2m-1}P(|\tilde Y_i|\geq x)dx \leq Cm\int x^{2m-1}\exp(-Cx^2)dx \\
& \leq 2CmC^{2m}m!\leq Cm^{3/2}C^mm^{m}.
\end{align}
Thus, applying Stirling's formula $m!\sim (2\pi m)^{1/2}(m/e)^m$ one more time yields
\[
E|P_{ii}\tilde Y_i^2|^m \leq Cm^{3/2}(CP_{ii})^mm^m\leq m!C^{m-2}P_{ii}^2/2, \text{ for } m\geq 2.
\]
By Bernstein's inequality (see \citeauthor{vandevaart1996},  \citeyear{vandevaart1996}), and noting that $\sum_{i=1}^n P_{ii}^2 \leq k $, we obtain that for any $x>0$,
\begin{align*}
& P\left(I_1(\alpha)\geq x^2 \right) = P\Big(\sum_{i=1}^nP_{ii}\tilde Y_i^2\geq kx^2 \Big) \leq 2\exp\Big(-\frac{k^2x^4}{2(\sum_{i=1}^nP_{ii}^2 + Ckx^2)}\Big)
%&\leq 2\exp\Big(-\frac{kx^4}{2(1+Cx^2)}\Big)
\leq 2\exp(-Ckx^2).
\end{align*}
Taking $x= \sqrt{\gamma_n\log p}$ with any $\gamma_n \rightarrow \infty$ we have
 \[
 P\big(I_1(\alpha)\geq \gamma_n\log p\big)\leq C\exp(-C\gamma_n|\alpha|\log p).
 \]
Since ${p\choose k} \leq (pe/k)^k$, it follows from the above inequality that
\begin{align*}
& P\big(\sup_{|\alpha|\leq K}I_1(\alpha) \geq \gamma_n\log p\big)\leq \sum_{|\alpha|\leq K}C\exp(-C|\alpha|\gamma_n\log p)\\
& \leq \sum_{k=1}^K{p\choose k}C\exp(-Ck\gamma_n\log p)\leq \sum_{k=1}^K(pe/k)^kC\exp(-Ck\gamma_n\log p)\rightarrow 0.
\end{align*}
Thus,
\begin{align}\label{e040}
\sup_{|\alpha|\leq K}I_1(\alpha) = O_p(\log p).
\end{align}

Now,  we consider $I_2(\alpha)$. Since $(Y_1, \cdots, Y_n)$ and $(Y_1^*, \cdots, Y_n^*)$ are independent copies,
by  the moment inequality (\ref{e057}) we have
\begin{align*}
&\sum_{i\neq j}|P_{ij}|^mE[|\tilde Y_i \tilde Y_j^*|^m] %=  \sum_{i\neq j}|P_{ij}|^mE[|\tilde Y_i|] E[|\tilde Y_j|^m]
\leq Cm^3C^mm^mP_{ij}^2\leq Cm^{5/2}C^mm!P_{ij}^2\leq m!C^{m-2}P_{ij}^2/2,
\end{align*}
for all $m\geq 2$ and $i\neq j \in\{1,\cdots, n\}$.
Thus, in view of (\ref{e012}), and by Bernstein's inequality and $\sum_{i\neq j}P_{ij}^2\leq k$,
\begin{align*}
 &P(|I_2(\alpha)| \geq x) \leq CP\big(\frac{1}{k}|\sum_{i\neq j}P_{ij}\tilde Y_i\tilde Y_j^*| \geq C^{-1}x \big)\\
  &\leq C\exp\Big(-\frac{1}{2C}\frac{k^2x^2}{\sum_{i\neq j}P_{ij}^2 + Cxk}\Big) \leq C\exp(-Ckx).
\end{align*}
Taking $x = \gamma_n\log p$ and using same argument as for $I_1(\alpha)$, we obtain that
\begin{align}\label{e041}
\sup_{|\alpha|\leq K}I_2(\alpha) = O_p(\log p).
\end{align}
Hence, the second result of Lemma \ref{P2} follows immediately from  (\ref{e040}) and (\ref{e041}).
\end{proof}

\section{Proof of Proposition \ref{T1}}

\begin{proof}
First consider non-Gaussian errors.
Define the event
\[\mathcal{E}_{n} = \{\sup_{|\alpha|\leq K}\{|\alpha|^{-1/2}\|\hbbeta^*(\alpha) - \bbeta^*(\alpha)\|_2\} \leq L_n\gamma_n\sqrt{(\log p)/n}\},\]
where $\gamma_n$ is some slowly diverging sequence and  $L_n$ is defined in Proposition \ref{P1}. The key of the proof is the following inequality
\begin{align}\label{e084}
\nonumber&P\Big(\sup_{|\alpha|\leq K}\frac{1}{|\alpha|}|\ell_n(\hbbeta(\alpha)) - \ell_n(\bbeta^*(\alpha))|\geq t \big|  \mathcal{E}_n\Big)\\
&\leq P\Big(\sup_{|\alpha|\leq K}\frac{1}{|\alpha|}|\ell_n(\hbbeta(\alpha)) - \ell_n(\bbeta^*(\alpha))|\geq t\big|  \mathcal{E}_{n}\Big) + P(\mathcal{E}_{n}^c).
\end{align}
Since $P(\mathcal{E}_{n}^c)=o(1)$ by Proposition \ref{P1},  we only need to consider the first term above.

For each given model $\alpha$, by definitions of $\hbbeta^*(\alpha)$ and $\bbeta^*(\alpha)$, we obtain that $\ell_n(\hbbeta^*(\alpha)) \geq \ell_n(\bbeta^*(\alpha))$ and
$
E\big[\ell_n(\bbeta^*(\alpha))-\ell_n(\hbbeta^*(\alpha))\big]  = I(\hbbeta^*(\alpha)) - I(\bbeta^*(\alpha)) \geq 0,
$
where $\hbbeta^*(\alpha)$ should be understood as a parameter of $E[\ell_n(\bbeta)]$.
Thus, conditioning on the event $\mathcal{E}_{n}$, with $N_n \equiv \gamma_nL_n\sqrt{|\alpha|(\log p)/n}$,
\begin{align*}
0\leq \ell_n(\hbbeta^*(\alpha)) - \ell_n(\bbeta^*(\alpha))\leq &\big(\ell_n(\hbbeta^*(\alpha)) -E[\ell_n(\hbbeta^*(\alpha))] \big) \\
&- \big(\ell_n(\bbeta^*(\alpha))-E[\ell_n(\bbeta^*(\alpha))] \big) \leq nZ_{\alpha}\big(N_n\big).
\end{align*}
In view of Lemma \ref{L2}, we have that, conditioning on $\mathcal{E}_n$,
\begin{align*}
\sup_{|\alpha|\leq K}\frac{1}{|\alpha|}\Big(\ell_n(\hbbeta^*(\alpha)) - \ell_n(\bbeta^*(\alpha))\Big) \leq O_p\big(L_n^2(\log p)\big).
\end{align*}
This, together with (\ref{e084}), completes the proof.

Now, we consider Gaussian errors. First, note that for a given model $\alpha$, we have
\begin{equation}\label{e096}
\bX_{\alpha}\t(\bX\bbeta_0 - \bX\bbeta^*(\alpha))=0.
\end{equation}
 So, using (\ref{e096}), and by direct calculations, we have $\hbbeta^*(\alpha)=\bbeta^*(\alpha)+(\bX_{\alpha}\t\bX_{\alpha})^{-1}\bX_{\alpha}\t\bW$ and
\begin{align*}
&\ell_n(\hbbeta^*(\alpha))%& = \bY\t\bX\hbbeta^*(\alpha) - \frac{1}{2}(\hbbeta^*(\alpha))\t\bX\t\bX\hbbeta^*(\alpha)\\
%&=
=\frac{1}{2}(\bbeta^*(\alpha))\t\bX\t\bX\bbeta^*(\alpha)+\bW\t\bX\bbeta^*(\alpha)+ \frac{1}{2}\bW\t\bX_{\alpha}(\bX_{\alpha}\t\bX_{\alpha})^{-1}\bX_{\alpha}\t\bW,\\
%\end{align*}
%On the other hand, by (\ref{e096}) and similar calculations we have
%\begin{align*}
&\ell_n(\bbeta^*(\alpha))= \bW\t\bX\bbeta^*(\alpha)+\frac{1}{2}(\bbeta^*(\alpha))\t\bX\t\bX\bbeta^*(\alpha).
\end{align*}
Combining the above two equations yields
$$
2(\ell_n(\hbbeta^*(\alpha)) - \ell_n(\bbeta^*(\alpha))) = \bW\t\bX_{\alpha}(\bX_{\alpha}\t\bX_{\alpha})^{-1}\bX_{\alpha}\t\bW,
$$
where the right-hand side term is $\chi^2_{|\alpha|}$ distributed. This follows that for any $t>0$,
\begin{align*}
P\Big(2(\ell_n(\hbbeta^*(\alpha)) - \ell_n(\bbeta^*(\alpha)))\geq |\alpha|t\Big)\leq C\exp(-C|\alpha|t).
\end{align*}
Using same argument as that for (\ref{e015}) completes the proof.
%\[
%\sup_{|\alpha|\leq K}\frac{1}{|\alpha|}(\ell_n(\hbbeta^*(\alpha)) - \ell_n(\bbeta^*(\alpha)) =O_p(\log p).
%\]

\end{proof}

\section{Proof of Proposition \ref{T5}}

\begin{proof}
By direct calculations,
$$
\ell_n(\bbeta^*(\alpha)) - E[\ell_n(\bbeta^*(\alpha))] = \bW\t\bX\bbeta^*(\alpha).
$$

If $W_i$'s are bounded, then by Hoeffding's inequality (see \citeauthor{BV11}, \citeyear{BV11})  we obtain that for any $t>0$,
\begin{align}\label{e060}
P\Big(\big|\bW\t\bX\bbeta^*(\alpha)\big|\geq t\Big)\leq C\exp\Big(-\frac{Ct^2}{\sum_{i=1}^n(\bx_i\t\bbeta^*(\alpha))^2}\Big).
\end{align}
Since $\sum_{i=1}^n(\bx_i\t\bbeta^*(\alpha))^2 = \bbeta^*(\alpha)\t\bX_{\alpha}\t\bX_{\alpha}\bbeta^*(\alpha) \leq Cn|\alpha| $,
if we take $t =|\alpha|\sqrt{n\gamma_n\log p}$ with $\gamma_n$ some slowly diverging sequence, then
\[
P\Big(\big|\bW\t\bX_{\alpha}\bbeta^*(\alpha)\big|\geq |\alpha|\sqrt{\gamma_nn\log p}\Big)\leq C\exp\big(-C\gamma_n|\alpha|\log p\big).
\]
Thus, using the same argument as that for (\ref{e015}) completes the proof.

For Gaussian errors and a given model $\alpha$,
$ \bW\t\bX_{\alpha}\bbeta^*(\alpha)\sim N\big(0, (\bbeta^*(\alpha))\t\bX_{\alpha}\t\bX_{\alpha}\bbeta^*(\alpha)\big).
$
Since $(\bbeta^*(\alpha))\t\bX_{\alpha}\t\bX_{\alpha}\bbeta^*(\alpha)\leq Cn|\alpha|$, it follows that
\[
P\Big(\ell_n(\bbeta^*(\alpha)) - E[\ell_n(\bbeta^*(\alpha))]\geq Ct\sqrt{|\alpha|n}\Big)\leq C\exp(-Ct^2).
\]
Taking $t=\sqrt{\gamma_n|\alpha|\log p}$, using the same argument as that for (\ref{e015}) completes the proof.

Finally, we consider unbounded errors. Similar to (\ref{e057}), we have for $m=2,3,\cdots$
\begin{align*}
 E[|\bx_i\t\bbeta^*(\alpha)W_i|^m] &\leq Cm(|\bx_i\t\bbeta^*(\alpha)|C)^{m}(m/2)!\leq (|\bx_i\t\bbeta^*(\alpha)|)^2(\|\bX\bbeta^*(\alpha)\|_{\infty}C)^{m-2}\frac{m!}{2}.
\end{align*}
Thus, an application of Bernstein's inequality yields that for any $t>0$,
\begin{align}\label{e035}
P\Big(|\bW\t\bX_{\alpha}\bbeta^*(\alpha)| \geq \sqrt{n}t\Big)\leq 2\exp\Big(-\frac{1}{2} \frac{nt^2}{C\|\bX_{\alpha}\t\bbeta^*(\alpha)\|_2^2 + C\sqrt{n}\|\bX_{\alpha}\bbeta^*(\alpha)\|_{\infty}t}\Big).
\end{align}
Note that $\|\bX_{\alpha}\t\bbeta^*(\alpha)\|_2^2 =O(|\alpha|n)$ and $\|\bX_{\alpha}\bbeta^*(\alpha)\|_\infty \leq \|\bX_{\alpha}\|_\infty\|\bbeta^*(\alpha)\|_\infty\leq C|\alpha|\max_{ij}|x_{ij}|$. Thus, $\|\bX_{\alpha}\bbeta^*(\alpha)\|_2^2/(\sqrt{n}\|\bX_{\alpha}\bbeta^*(\alpha)\|_{\infty})\geq n^{\tau}$. Taking $t =|\alpha|\sqrt{\gamma_n\log p}$, then if $K^2(\log p)/n^{2\tau}\rightarrow 0$, we have $\|\bX_{\alpha}\bbeta^*(\alpha)\|_2^2\gg \sqrt{n}\|\bX_{\alpha}\bbeta^*(\alpha)\|_{\infty}t$, and thus (\ref{e035}) becomes
\begin{align*}
P\Big(|\bW\t\bX_{\alpha}\bbeta^*(\alpha)| \geq |\alpha|\sqrt{\gamma_n n\log p}\Big)\leq 2\exp\big(-C\gamma_n|\alpha|\log p\big).
\end{align*}
Using a similar argument as that for (\ref{e015}) completes the proof.

\end{proof}

\end{document}